\documentclass[conference]{IEEEtran}
\IEEEoverridecommandlockouts
\usepackage{cite}
\usepackage{amsmath,amssymb,amsfonts}
\usepackage{algorithmic}
\usepackage{graphicx}
\usepackage{textcomp}
\usepackage{xcolor}
\usepackage{booktabs}

\usepackage{hyperref}
\usepackage{titlesec}
\usepackage{float}
\usepackage{orcidlink}
\usepackage{subcaption} 
\titlespacing*{\subsection}{0pt}{4pt}{3pt} 

\def\BibTeX{{\rm B\kern-.05em{\sc i\kern-.025em b}\kern-.08em
    T\kern-.1667em\lower.7ex\hbox{E}\kern-.125emX}}

\begin{document}

\title{Securing LLMs in the Wild: Privacy and Security Challenges at the Edge\\
}

\author{
    \IEEEauthorblockN{
        Ren-Yi Huang\,\orcidlink{0009-0006-8341-5835},
        Mingchen Li\,\orcidlink{0000-0003-4553-9598},
        Dumindu Samaraweera\,\orcidlink{0000-0003-4097-5585},\textit{Senior Member, IEEE},
        and \\
        Morris Chang\,\orcidlink{0000-0002-0660-7191},\textit{Member, IEEE}
    }
    
    \thanks{
    Ren-Yi Hunag, Mingchen Li, and Morris Chang are with the Department of Electrical and Computer Engineering, University of South Florida. (e-mail: hr219@usf.edu; mingchenli@usf.edu; chang5@usf.edu).
    } 
    \thanks{
    Dumindu Samaraweera is with the Department of Mathematics, Embry-Riddle Aeronautical University. (e-mail: samarawg@erau.edu).
    }
}

\maketitle

\titlespacing*{\section}{0pt}{0.5\baselineskip}{0.3\baselineskip}
\titlespacing*{\subsection}{0pt}{0.3\baselineskip}{0.1\baselineskip}

\begin{abstract}
    Large Language Models (LLMs) are rapidly moving from isolated research environments to the "wild", where they are increasingly deployed on enterprise infrastructure, personal devices, and edge platforms. While cloud-based deployments provide scalable computational resources, they also raise significant security and privacy concerns related to data sovereignty, regulatory compliance, latency, and dependence on third-party providers. As a result, organizations are increasingly adopting edge and on-premise LLM deployments to gain greater control over sensitive data and operational workflows. This architectural shift, however, introduces a new class of security and privacy challenges. In particular, limited computational and memory resources require models to undergo aggressive optimizations, including quantization, pruning, model partitioning, and parameter-efficient adaptation, each of which can introduce new vulnerabilities and fundamentally reshape the threat landscape. We describe this tension as the \emph{Security–Efficiency Paradox}, in which mechanisms designed to improve deployment efficiency may inadvertently weaken model robustness, expose new attack surfaces, or increase privacy risks. In particular, we examine how model compression can affect safety alignment, how partitioned inference can enable information reconstruction attacks, and how continuous local adaptation may introduce privacy leakage and model drift. To systematically analyze these risks, we introduce a deployment-centric taxonomy organized around three fundamental architectural constraints: the Memory Wall, the Quadratic Wall, and the Compute Wall. We then derive a unified constraint model that quantifies when unsafe optimizations become unavoidable, linking each wall directly to specific attack surfaces. Building on this model, we propose the Secure Operational Efficiency Score (SOES), a holistic metric that balances task accuracy, jailbreak resistance, and privacy against energy, memory, and latency, enabling practitioners to select and configure edge LLMs under real-world hardware limits. We further present a practical decision procedure and targeted mitigations for each optimization-induced vulnerability. Together, these contributions move beyond isolated defenses toward a co-designed framework for evaluating security, privacy, and efficiency simultaneously, providing a foundation for securing the next generation of edge-native intelligent systems.
\end{abstract}


\section{Introduction} \label{introduction}
The integration of Large Language Models (LLMs) into production-ready software systems has fundamentally reshaped human-computer interaction and automated reasoning. However, the standard operating paradigm, whereby local client applications offload inference to centralized, multi-tenant cloud APIs, presents severe systemic vulnerabilities \cite{das2025security}. Enterprise data leakage, exposure of intellectual property, and regulatory non-compliance (e.g., GDPR \cite{GDPR2016}, HIPAA \cite{HIPAAPrivacyRule2002}, EU AI Act \cite{EUAIAct2024}, and the U.S. AI Action Plan \cite{USAIActionPlan2025}) serve as major deterrents to cloud-hosted AI. At the same time, emerging AI governance initiatives, such as Google's DeepMind Frontier Safety Framework \cite{DeepMindFSF2025}, emphasize that advanced AI systems must be evaluated not only for capability, but also for safety, security, and responsible deployment throughout their lifecycle. Hence, the balance between practical deployment, regulatory compliance, and robust security assurance is far from realized.

To address these challenges, an architectural paradigm shift is underway: deploying LLMs directly to the edge, including local workstations, corporate on-premise servers, and tactical edge hardware \cite{kristiani2026deploying, friha2024llm}. By keeping sensitive data and inference local, edge deployment offers a promising path toward improved privacy, reduced latency, and greater operational control. Yet, moving models "into the wild" forces an immediate collision with severe hardware limitations. LLM deployment is fundamentally constrained by three distinct architectural bottlenecks:

\begin{figure*}
    \centerline{\includegraphics[width=42pc]{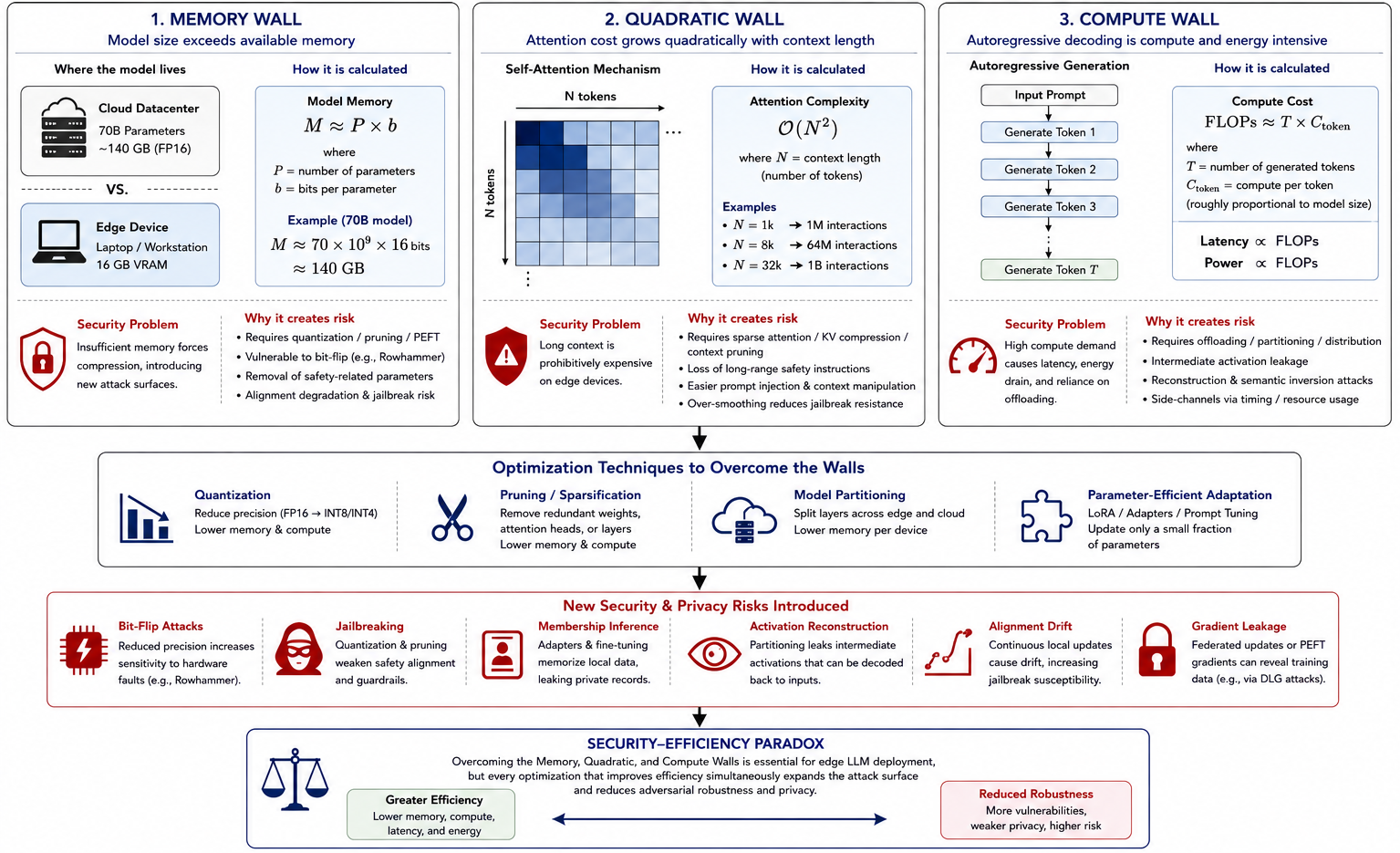}}
    \caption{Architectural bottlenecks and security tradeoffs in edge LLM deployment. To overcome the limitations of three fundamental barriers: the Memory Wall, Quadratic Wall, and Compute Wall, developers employ optimization techniques such as quantization, pruning, model partitioning, and parameter-efficient adaptation. While these approaches reduce memory, computational, and energy requirements, they simultaneously introduce new attack surfaces, creating a fundamental security–efficiency paradox for edge AI systems.}
    \label{fig:sec_eff_paradox}
    \vspace*{-5pt}
\end{figure*}

\begin{enumerate}
    \item The Memory Wall: Modern LLMs contain billions of parameters, requiring gigabytes (or even hundreds of gigabytes) of memory to store model weights and intermediate computations. Most consumer devices and edge platforms simply lack sufficient memory capacity, making it difficult to host large models locally.
    
    \item The Quadratic/Attention Wall: LLMs rely on self-attention mechanisms to understand relationships between tokens in a prompt. Unfortunately, the computational and memory requirements of this process grow rapidly as the context length increases (usually $O(N^2)$). As a result, supporting long documents, extensive conversations, or large retrieval-augmented contexts becomes increasingly expensive and often impractical on resource-constrained hardware \cite{kwon2023efficient}.
        
    \item The Compute Wall: Generating text requires continuous autoregressive decoding, where each new token is produced based on all previously generated tokens. This process demands substantial computational throughput (usually measured in floating-point operations per second, FLOPS) and energy consumption. On edge devices, limited processing power and battery constraints can lead to higher latency, reduced responsiveness, and shortened device lifetime.  
\end{enumerate}

To overcome these barriers, developers increasingly rely on model compression, quantization, and architectural partitioning techniques. While these approaches enable efficient deployment on resource-constrained hardware, they also introduce new security and privacy risks. This paper examines the unintended consequences of such optimizations and explores the fundamental tradeoff between system efficiency and adversarial robustness. Figure~\ref{fig:sec_eff_paradox} illustrates this security-efficiency paradox: the very optimizations that make edge deployment feasible also create new avenues for adversarial attacks, including jailbreaks, membership inference, and gradient leakage. To address these challenges, this paper makes the following contributions:

\begin{enumerate}
    \item We systematically analyze the Security-Efficiency Paradox in edge LLM deployment. We examine four representative optimization families (quantization, pruning, model partitioning, and parameter-efficient adaptation) and demonstrate how each technique, while enabling deployment on resource-constrained hardware, introduces distinct attack surfaces including jailbreak vulnerabilities, safety subnetwork fragmentation, input reconstruction attacks, and membership inference leakage.
    
    \item We derive a unified \emph{Three-Wall} Constraint Model that quantitatively characterizes the Memory, Quadratic, and Compute Walls governing edge LLM feasibility. This model defines when unsafe optimizations become unavoidable, linking each hardware constraint directly to specific security risks and establishing a principled foundation for security-aware model selection.

    \item We propose the \emph{Secure Operational Efficiency Score (SOES)}, a holistic metric that jointly captures task accuracy, jailbreak resistance, and privacy preservation against energy consumption, VRAM footprint, and inference latency. This enables practitioners to systematically compare and select edge-optimized models under real-world hardware constraints.

    \item We present a practical four-stage deployment pipeline that integrates hard-wall filtering, mixed-precision quantization, noise-calibrated split inference, and SOES-based optimization to guide secure model selection and configuration.

    \item We conduct a comparative empirical evaluation of a series of representative edge-friendly LLMs across multiple quantization configurations, validating the Security-Efficiency Paradox and demonstrating the utility of SOES for guiding deployment decisions under resource constraints.
\end{enumerate}

The remainder of this paper is structured as follows. Section II examines the Security-Efficiency Paradox in detail, and the Section III presents our Three-Wall Constraint Model, quantifying the Memory, Quadratic, and Compute Walls and deriving a unified feasibility criterion for secure edge deployment. Section IV introduces the Secure Operational Efficiency Score (SOES) and describes our four-stage deployment pipeline. Section V presents our comparative experimental evaluation of edge-friendly LLMs, validating the proposed metrics and demonstrating practical model selection under hardware constraints. Finally, Section VI concludes the paper and discusses directions for future work.

\section{The Security-Efficiency Paradox at the Edge} \label{sec:sec_eff_paradox}
\begin{figure*}
    \centerline{\includegraphics[width=42pc]{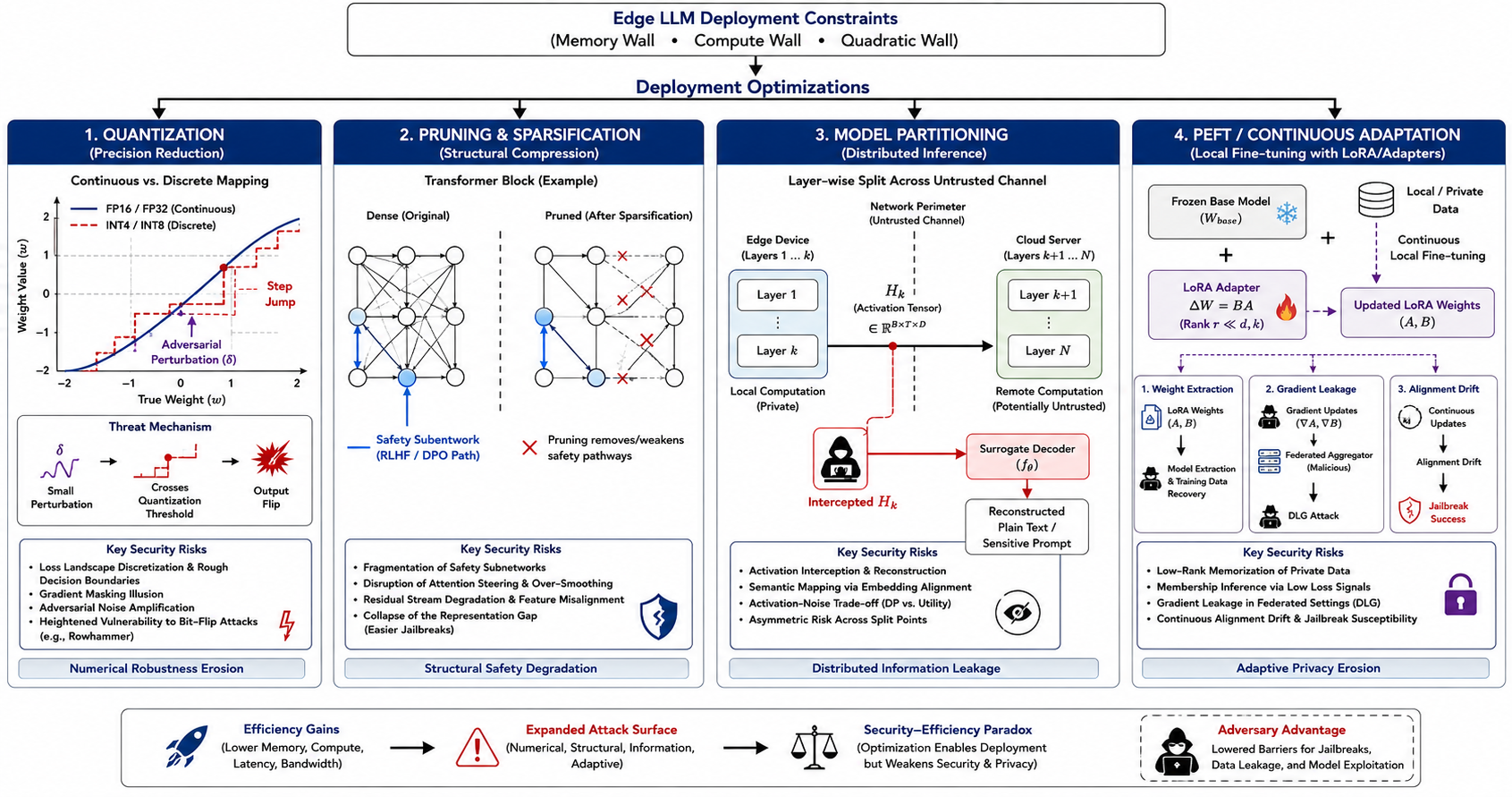}}
    \caption{Taxonomy of decentralized LLM deployment trade-offs. The diagram contrasts how static model optimizations and runtime operational frameworks compromise structural safety components, converting hardware efficiency features into active adversarial attack vectors. While these techniques enable deployment on resource-constrained hardware, each introduces a distinct attack surface. Quantization weakens numerical robustness, pruning degrades safety-alignment subnetworks, partitioning exposes intermediate activations to reconstruction attacks, and continuous local adaptation increases susceptibility to privacy leakage and alignment drift. Together, these optimizations illustrate the fundamental tension between deployment efficiency and trustworthy AI operation.}
    \label{fig:optimization_str}
    \vspace*{-5pt}
\end{figure*}

To understand the security--efficiency paradox, we must first examine how aggressive optimization techniques operate under the constraints of resource-limited deployment environments. Building upon our prior work \cite{huang2025advancing}, this section investigates how these techniques reshape the security landscape by introducing new attack vectors and privacy risks, as illustrated in Fig.~\ref{fig:optimization_str}. To provide a structured discussion, we classify existing optimization approaches into four representative families: quantization, pruning, model partitioning, and parameter-efficient adaptation. Each family is examined in terms of its underlying optimization mechanism, deployment benefits, and associated security implications.

\subsection{\textbf{Quantization: Precision Reduction and Its Security Implications}}
To overcome the Memory Wall and reduce the Compute Wall, edge deployments commonly employ quantization. Quantization reduces the numerical precision of model weights and activations (e.g., from FP32 to INT8 or INT4) to decrease memory footprint and accelerate inference on resource-constrained hardware \cite{yao2022zeroquant}. While these gains are critical for edge deployment, quantization introduces several underappreciated security consequences.

\begin{itemize}
    \item \textit{Loss Landscape Discretization and Rough Decision Boundaries:}\\
    In a standard full-precision neural network, the loss landscape is smooth and continuous, allowing optimization algorithms (and adversaries) to navigate the parameter space using precise gradient information. To reduce the memory footprint and computational cost of edge deployment, quantization represents weights ($W$) and activations ($X$) using low-bit integer values instead of high-precision floating-point representations. Specifically, a floating-point tensor $X \in \mathbb{R}$ is mapped to a lower-precision integer tensor $X_q \in \mathbb{Z}$ through a scaling factor $S$ and a zero-point offset $Z$:
    
    $$
    X_q = \text{clip}\left( \left\lfloor \frac{X}{S} \right\rceil + Z, q_{min}, q_{max} \right)
    $$
    
    This quantization process fundamentally transforms the optimization landscape by replacing a continuous parameter space with a discrete one. The clipping and rounding operations introduce discontinuities that reshape the model's decision boundaries from smooth geometric manifolds into piecewise "staircase" structures. Consequently, input samples that lie near an original decision boundary can be shifted across these discrete thresholds by only small, often imperceptible perturbations, increasing the model's sensitivity to distribution shifts and adversarial examples \cite{egashira2024exploiting}. As a result, the low-precision representations introduced to overcome the Memory Wall reduce the stability of the model's decision boundaries, making adversarial manipulations and safety-alignment failures more likely.
  
    \vspace{5pt}
    \item \textit{The Illusion of Gradient Masking:}\\
    Attackers often use gradient-based methods (like the Fast Gradient Sign Method~\cite{goodfellow2014explaining}) to calculate exactly how to alter an input to trick a model. In a heavily quantized network, the derivative of a discrete rounding function is zero almost everywhere. When an attacker tries to compute standard gradients to find a vulnerability, the gradients appear to vanish or become highly chaotic. This phenomenon is known as gradient masking~\cite{galloway2017attacking}. While gradient masking makes the model appear resistant to simple gradient attacks, it provides a false sense of security. Attackers easily bypass this by utilizing Expectation over Transformation (EOT) ~\cite{athalye2018synthesizing}, black-box optimization algorithms, or transfer attacks (generating the attack on a smooth, full-precision surrogate model and deploying it against the quantized model)~\cite{yang2024quantization}. Because the surrogate's smooth gradients approximate the quantized model's general direction, transfer attacks often achieve a higher success rate against quantized targets than dense ones.

    \vspace{5pt}
    \item \textit{Amplification of Adversarial Noise:}\\
    Every layer of a neural network processes the output of the previous layer. In a quantized network, every layer introduces a small amount of quantization noise (rounding error), denoted as $\mathcal{E}$:
    $$\hat{X} = X + \mathcal{E}$$
    
    When a model runs normally, these small errors ($\mathcal{E}_1, \mathcal{E}_2, \dots$) are bounded and usually cancel out. However, when an adversary intentionally injects a structured perturbation ($\delta$) into the input, the network's internal activations change.

    In an unquantized network, safety alignments or robust layers might dampen minor adversarial noise before it reaches the output. In a quantized network, the adversarial perturbation $\delta$ interacts constructively with the quantization error $\mathcal{E}$ at each layer. As the forward pass progresses through dozens of layers, this error accumulates and magnifies. By the final layer, a perturbation that an FP16 model would have ignored completely flips the output probability, resulting in a successful jailbreak or misclassification \cite{zahran2025jailbreaking, lee2025quantization}. This demonstrates how quantization employed to satisfy the Memory and Compute Walls can amplify small adversarial perturbations into safety-critical failures during edge inference.

    \vspace{5pt}
    \item \textit{Heightened Vulnerability to Bit-Flip Attacks:}\\
    Quantization drastically reduces the number of bits representing a single weight. In an INT4 model, each parameter is assigned only 4 bits of memory. In hardware-level security, techniques like Rowhammer ~\cite{kim2014flipping} allow an attacker to induce electrical disturbances in neighboring memory cells on a DRAM chip, flipping a bit from a 0 to a 1 or vice versa without needing software permissions.

    For instance, in an FP32 model, flipping a single bit in a mantissa has a negligible effect on the overall parameter value. In an INT4 model, flipping just one bit can alter the parameter value by 50\% or flip its sign completely~\cite{rakin2019bit}. If an attacker targets a highly significant bit within a critical safety-alignment layer or an attention head responsible for filtering toxic tokens, a single bit-flip can completely disable the model's safety guardrails. Consequently, reducing numerical precision to satisfy edge hardware constraints also lowers the effort required for hardware-level fault injection attacks that compromise model integrity and safety. 
    
\end{itemize}

Ultimately, quantization represents the first manifestation of the Security-Efficiency Paradox. While it enables edge deployment by overcoming the Memory Wall and alleviating the Compute Wall, the resulting loss of numerical precision fundamentally weakens the model's robustness, expanding the attack surface for adversarial manipulation, fault-injection attacks, and safety-alignment failures.

\subsection{\textbf{Pruning: Sparsity, Robustness, and Attack Surface Expansion}}
To reduce computational complexity imposed by the Compute Wall while simultaneously lowering memory requirements, pruning removes redundant weights, attention heads, or entire layers to produce sparse or structurally compact models. While pruning can reduce compute and communication overhead in federated and edge settings, it fundamentally alters the geometry of the model's decision surface in ways that carry security implications. To understand how weight pruning and sparsification inadvertently degrade an LLM's safety filters, we must examine how safety alignment (such as RLHF ~\cite{christiano2017deep, stiennon2020learning}, DPO ~\cite{rafailov2023direct}, or system-prompt enforcement) is structurally represented inside a Transformer network~\cite{qi2025safety}.

\begin{itemize}
    \item \textit{The Fragmentation of Safety Subnetworks:}\\
    Deep learning research suggests that within a massive neural network, specialized capabilities are often localized in specific, sparse pathways known as subnetworks \cite{frankle2018lottery}. Safety alignment is rarely distributed evenly across every single parameter; instead, it typically relies on highly specialized attention heads and residual stream channels that learn to identify toxic prompts and suppress malicious completions. When we apply unstructured or semi-structured pruning, the compression algorithm removes weights based on a purely statistical heuristic, usually magnitude-based pruning (removing weights closest to zero \cite{sun2024simple} or first-order gradient methods (removing weights that contribute least to the baseline language modeling loss) \cite{ma2023llm}.

    Because safety alignment represents a tiny fraction of the overall training objective compared to general language modeling, the pruning algorithm treats safety-related pathways as "noise" or "low-importance redundancy." By clipping these low-magnitude but highly specialized weights, the algorithm inadvertently fragments the safety subnetwork, while leaving the core capability to generate coherent text completely intact~\cite{wei2024assessing}. Consequently, pruning strategies introduced to satisfy the Memory and Compute Walls may preserve general task performance while selectively weakening safety-critical subnetworks, increasing the likelihood of unsafe model behavior.

    \vspace{5pt}
    \item \textit{Disruption of Attention Steering and Over-Smoothing:}\\
    In a Transformer, multi-head attention (MHA)~\cite{vaswani2017attention} allows the model to maintain awareness of system instructions, safety policies, and other contextual constraints throughout a conversation. Certain attention pathways play an important role in reinforcing these constraints by directing the model toward safe and policy-compliant responses. However, sparsification often removes entire attention heads (structured pruning) or heavily sparsifies the Key and Value projection matrices ($W_k, W_v$) to reduce the quadratic KV cache bottleneck~\cite{michel2019sixteen, chang2024palu}.

    When these attention projection matrices are sparsified, the model's ability to maintain long-range context dependency degrades. As a result, the model may place greater emphasis on recently observed inputs while losing sensitivity to earlier safety instructions. This phenomenon, often referred to as \textit{attention over-smoothing}, can create opportunities for adversaries to craft prompts that override or bypass built-in safeguards~\cite{ma2026efficiency}. For example, a carefully designed jailbreak prompt embedded near the end of a lengthy interaction may receive disproportionate attention, causing the model to ignore safety directives provided earlier in the context window. Consequently, deployment-driven optimizations intended to improve efficiency may inadvertently increase the susceptibility of LLMs to prompt-injection and jailbreak attacks. This illustrates how sparsification techniques designed to alleviate the Quadratic and Compute Walls can inadvertently degrade long-range safety enforcement, creating new opportunities for prompt-injection and jailbreak attacks.

    \vspace{5pt}
    \item \textit{Residual Stream Perturbation and Representation Collapse:}\\
     Modern LLMs process information through a shared "residual stream," where each transformer layer reads from and writes to a common high-dimensional representation. Safety-aligned models embed refusal behaviors/responses (e.g., "I cannot fulfill this request") and alignment constraints within this residual space by learning feature directions that suppress unsafe generations while preserving benign capabilities. However, pruning fundamentally perturbs the geometry of this representation. By removing neurons and compressing intermediate feed-forward network (FFN) layers, pruning alters how semantic features are encoded and combined, causing safety-related feature vectors to become misaligned with the harmful representations they were originally trained to suppress~\cite{hasan2024pruning}.

    As the degree of sparsification increases, these local perturbations propagate throughout the network and reduce the separability of the latent representation space. In a dense model, benign and adversarial prompts occupy distinguishable regions of the embedding manifold, allowing safety mechanisms to reliably recognize malicious intent and trigger refusal behaviors. Pruning compresses this representation space, shrinking the geometric margin between benign and adversarial prompts. Consequently, adversarial jailbreak prompts become increasingly similar to legitimate requests in the model's internal feature space, reducing the effectiveness of safety classifiers and alignment mechanisms. The result is a model that largely preserves its general reasoning capabilities while becoming significantly more susceptible to jailbreak attacks and other prompt-based adversarial manipulations.
    
\end{itemize}
Ultimately, pruning represents another manifestation of the Security-Efficiency Paradox. While it alleviates the Memory and Compute Walls by removing redundant parameters and reducing inference complexity, it also fragments safety-critical subnetworks, disrupts attention mechanisms, and compresses internal representations, thereby increasing susceptibility to jailbreak attacks, adversarial manipulation, and alignment failures.

\subsection{\textbf{Model Partitioning: Distributed Inference and Information Reconstruction Attacks}}
Model partitioning splits a large model across multiple devices or nodes, vertically (pipeline parallelism, layer-wise splitting) or horizontally (tensor parallelism), to overcome the Memory Wall and enable inference on hardware incapable of hosting the full model~\cite{zhang2024edgeshard}. This architectural pattern introduces a qualitatively new threat surface absent in monolithic cloud deployments \cite{he2020attacking}. When an LLM is split across devices, the network transmissions are not raw text, but dense mathematical vectors. While this appears to act as a natural layer of obfuscation, it actually introduces a highly exploitable structural vulnerability.

\begin{itemize}
    \item \textit{Reversing the Feed-Forward Pipeline:}\\
    In a partitioned Transformer inference setup, the local edge device computes the first $k$ layers of the network and transmits the resulting activation tensor $H_k \in \mathbb{R}^{B \times T \times D}$ (where $B$ is batch size, $T$ is token length, and $D$ is the hidden dimension) over an untrusted network to a more powerful server.The server is intended to compute the remaining layers ($k+1$ to $N$). However, an adversary intercepting $H_k$ (or a malicious server operator) can treat the edge layers as a forward mathematical function that they want to invert.

    Because deep neural networks are designed to retain maximum semantic information as data flows deeper into the network, this $H_k$ acts as a dense lossy compression of the original input. An attacker can train a lightweight surrogate decoder model (often an MLP or a small transformer-based decoder) configured to reverse the process. By optimizing a reconstruction loss function (such as cross-entropy over the vocabulary space), the attacker can decode the continuous vector $H_k$ back into the exact sequence of discrete textual tokens originally entered by the user \cite{luo2025prompt}. Therefore, model partitioning adopted to overcome the Memory Wall transforms intermediate activations into a high-value attack surface, enabling reconstruction of sensitive user inputs from intercepted hidden representations.

    \vspace{5pt}
    \item \textit{Semantic Mapping:}\\
    One could assume that as a prompt travels through deeper layers, the individual tokens become so thoroughly mixed and contextualized that the original words disappear. However, language model embeddings operate on semantic manifolds where structural relationships are preserved geometrically. For instance, the word vector for "Apple" and "iPhone" will maintain a distinct geometric distance and cluster relationship within $H_k$, even after passing through several attention blocks.

    However, the downside is that an attacker does not even need to train an expensive decoder from scratch. By using Embedding Space Alignment, they can map the intercepted activation distributions to a known, publicly available reference model (like an unslotted base version of Llama or Mistral). Once the coordinate geometry of the hidden space is aligned via linear regression, the attacker can use the public model's unembedding layer ($W_{te}$) to instantly project the hidden states back into readable tokens, bypassing any perimeter security completely \cite{chen2025algen}. As a result, partitioning not only exposes hidden representations but also enables efficient semantic reconstruction attacks that compromise user privacy without requiring access to the original proprietary model.

    \vspace{5pt}
    \item \textit{The Activation-Noise Trade-off:}\\
    In order to prevent interception and reconstruction attacks, developers usually inject noise into the activations before transmission, commonly enforcing Differential Privacy (DP)~\cite{dwork2006calibrating} or adding Gaussian noise to mask individual token signatures:
    $$\tilde{H}_k = H_k + \mathcal{N}(0, \sigma^2)$$

    In practice, this creates a severe operational bottleneck. Because LLM performance is incredibly sensitive to the precise coordinates of its hidden states, even minor amounts of noise ($\sigma$) break the attention arithmetic in the subsequent layers ($k+1$ to $N$). If the noise is strong enough to hide the semantic identity of the tokens from an attacker, it is usually strong enough to cause the remaining cloud-side layers to generate incoherent gibberish which typically lacks logical structure (halting task execution or causing catastrophic perplexity spikes). This exposes a fundamental privacy-utility trade-off in partitioned inference, where protecting intermediate activations often directly conflicts with maintaining acceptable inference quality on edge devices.

    \vspace{5pt}
    \item \textit{Asymmetric Vulnerability Across Split Points:}\\
    Another important aspect is where the model split occurs. The level of risk is heavily dependent on where the model is cut. If the model is cut near the early layers (e.g., Layer 2 or 3), the activations behave almost exactly like token embeddings, making reconstruction trivial. If cut near the very end (e.g., Layer 30 of 32), reconstruction is difficult, but the edge device must handle 90\%+ of the computation, rendering the cloud offloading useless.

    In reality, systems often choose a middle ground (e.g., splitting at Layer 12 or 16) to balance hardware capability. This midpoint represents the worst of both worlds from a security perspective: the compute load is shared, but the activation vectors have formed dense, stable contextual abstractions. At this stage, things like structural syntactic patterns, PII, and proprietary code formatting are deeply etched into the hidden trajectories, giving a reconstruction attack a wealth of structural context to piece together complete sensitive documents. Thus, selecting the partition boundary becomes a joint optimization problem governed by both the Memory Wall and privacy considerations, where improving computational efficiency can substantially increase reconstruction risk.
\end{itemize}

Ultimately, model partitioning represents another manifestation of the Security-Efficiency Paradox. While it overcomes the Memory Wall by distributing computation across heterogeneous edge and cloud resources, it simultaneously exposes intermediate activations that become valuable attack surfaces for input reconstruction, semantic inference, and privacy leakage.

\subsection{\textbf{Parameter-Efficient Adaptation: Continuous Local Fine-Tuning and Privacy Leakage}}
To overcome both the Compute Wall and Memory Wall during local adaptation, parameter-efficient fine-tuning (PEFT) methods \cite{dettmers2023qlora}, including Low-Rank Adaptation (LoRA) \cite{hu2022lora}, prompt tuning, prefix tuning, and adapter modules, enable LLMs to be adapted to downstream tasks by updating a small fraction of parameters. While PEFT bypasses the Compute and Memory Walls by freezing the base model and updating only a tiny fraction of parameters, it introduces severe security and privacy vulnerabilities. Because these adapters are explicitly trained to compress and memorize localized, domain-specific information, they transform the edge architecture into a highly targeted attack surface.

\begin{itemize}
    \item \textit{The Low-Rank Memorization Matrix:}\\
    LoRA modifies the weight update $\Delta W$ of a dense attention or feed-forward layer by decomposing it into two low-rank matrices, $A$ and $B$, where $W_{new} = W_{base} + B \cdot A$:
    $$\Delta W = B \cdot A$$

    where: $$ B \in \mathbb{R}^{d \times r}, A \in \mathbb{R}^{r \times k}, \text{ and } r \ll \min(d, k)$$
    
    The rank $r$ is typically very small (e.g., $r=8$ or $r=16$). This extreme reduction in dimensionality forces the low-rank matrices to act as a highly condensed semantic funnel.

    Because the base model remains entirely frozen, the adapter matrices ($A$ and $B$) are forced to capture only what is novel about the local dataset (e.g., specific corporate IP, local user credentials, financial transactions, or may be PII). When an attacker obtains access to the edge device, they do not need to audit a massive 70-billion-parameter network. They can extract just the few megabytes containing the LoRA weights. Because of the mathematical isolation of $\Delta W$, an attacker can isolate, reverse-engineer, or query these lightweight parameters to reconstruct verbatim training sequences far more easily than they could from a globally fine-tuned dense model~\cite{arazzi2026lora}. Consequently, PEFT methods introduced to overcome the Compute and Memory Walls concentrate domain-specific knowledge into compact adapter parameters, substantially increasing the risk of parameter extraction and sensitive data reconstruction.

    \vspace{5pt}
    \item \textit{Amplified Vulnerability to Membership Inference Attacks:}\\
    Membership Inference Attacks (MIAs) \cite{shokri2017membership, carlini2021extracting, carlini2022quantifying} aim to determine whether a specific data record (e.g., a patient's medical history, proprietary code snippet, or financial transaction) was included in the model's fine-tuning dataset. In the context of PEFT-based edge adaptation, MIAs exploit the fact that the adapted model assigns systematically higher confidence (or lower loss) to training samples than to unseen data. Formally, let $x = (x_1, x_2, \dots, x_T)$ denote an input token sequence, and let the adapted model be parameterized by the frozen base weights $W_{\text{base}}$ and the learned low-rank adapter update $\Delta W = B \cdot A$. The membership decision can then be expressed as:
    $$
    \mathcal{P}_{\text{member}} =
    \mathbb{I}\left[
    -\sum_{t=1}^{T} \log P\big(x_t \mid x_{<t};\, W_{\text{base}} + B \cdot A\big)
    < \tau
    \right]
    $$
    
    where: $P(x_t \mid x_{<t}; \cdot)$ denotes the conditional token probability at position $t$ given prior context, $x_{<t}$ represents all tokens preceding position $t$, $W_{\text{base}} + B \cdot A$ is the PEFT-adapted model composed of a frozen backbone and a low-rank update, $\sum_{t=1}^{T} -\log P(\cdot)$ is the sequence-level negative log-likelihood (NLL), $\tau$ is a detection threshold calibrated to distinguish member from non-member samples, and $\mathbb{I}[\cdot]$ is the indicator function.
    
    In edge-deployed PEFT settings, this vulnerability is further amplified due to the inherently small and highly specialized nature of local fine-tuning datasets. Such settings induce rapid and localized overfitting in the adapter parameters $A$ and $B$, causing the model to assign disproportionately high confidence to memorized or frequently observed sequences. Consequently, when an adversary queries the system with a candidate record, an anomalously low loss (or equivalently, high likelihood) serves as a statistical fingerprint of data inclusion within the private fine-tuning set, thereby enabling successful membership inference under even limited query access. Therefore, efficient local adaptation on small edge datasets unintentionally amplifies memorization effects, increasing susceptibility to membership inference attacks against sensitive training records.

    \vspace{5pt}
    \item \textit{Gradient Leakage in Local Federated Aggregations:}\\
    In many advanced edge applications, separate on-premise nodes use Federated Learning (FL) \cite{flsec_ren-yi} to collaboratively update an LLM without sharing raw data. Nodes compute LoRA gradient updates ($\nabla A, \nabla B$) locally and transmit only these weight updates to a central aggregator \cite{bai2024federated}.

    However, this architecture is highly vulnerable to Deep Leakage from Gradients (DLG) \cite{zhu2019deep}. A malicious central aggregator or a man-in-the-middle attacker can intercept the PEFT gradient updates. Because the gradients are highly focused on the low-rank adjustments of the safety and semantic steering paths, the attacker can initialize dummy inputs and dummy labels, feed them into a surrogate network, and iteratively optimize them to match the intercepted gradients:
    $$\min_{X^*} \|\nabla W_{\text{dummy}}(X^*) - \nabla W_{\text{local}}\|^2_2$$

    where $\nabla \Theta(X; W_{\text{base}}, A, B)$ denotes the gradients computed on the dummy input $X$ through the forward and backward passes of the PEFT-adapted model. Because the parameter space of a LoRA update is so small, this optimization problem converges significantly faster than traditional dense gradient inversion, allowing an attacker to reconstruct the exact private text tokens processed by the edge node. Although federated PEFT reduces communication and computational overhead, the resulting compact gradient updates create an efficient attack surface for gradient inversion and private data reconstruction.

    \vspace{5pt}
    \item \textit{Continuous Alignment Drift:}\\
    As an edge model undergoes continuous localized updates to adapt to local jargon or operational environments, it experiences a phenomenon known as Alignment Drift (a specialized form of catastrophic forgetting where the model gradually loses the behaviors, values, or safety constraints that were originally taught to it)~\cite{das2025alignguard}.

    During the initial alignment phase (RLHF/DPO), the base model is trained to reject harmful prompts (e.g., malware generation or dangerous data extraction). However, the continuous, localized updates to the LoRA parameters alter the mathematical trajectories in the network's residual stream. Over time, the newly learned local weights begin to mask or overwrite the frozen base model's safety activation pathways. Ultimately, continuous PEFT updates intended to maintain efficient on-device adaptation gradually erode the model's original safety alignment, increasing long-term vulnerability to jailbreak, prompt injection, and policy-evasion attacks.

\end{itemize}
Ultimately, parameter-efficient adaptation represents another manifestation of the Security-Efficiency Paradox. While techniques such as LoRA and other PEFT methods overcome the Memory and Compute Walls by enabling efficient on-device fine-tuning, they also concentrate task-specific knowledge into compact adapter parameters, increasing the risks of membership inference, gradient leakage, and long-term safety alignment drift.

\begin{figure}[H]
    \centerline{\includegraphics[width=\columnwidth]{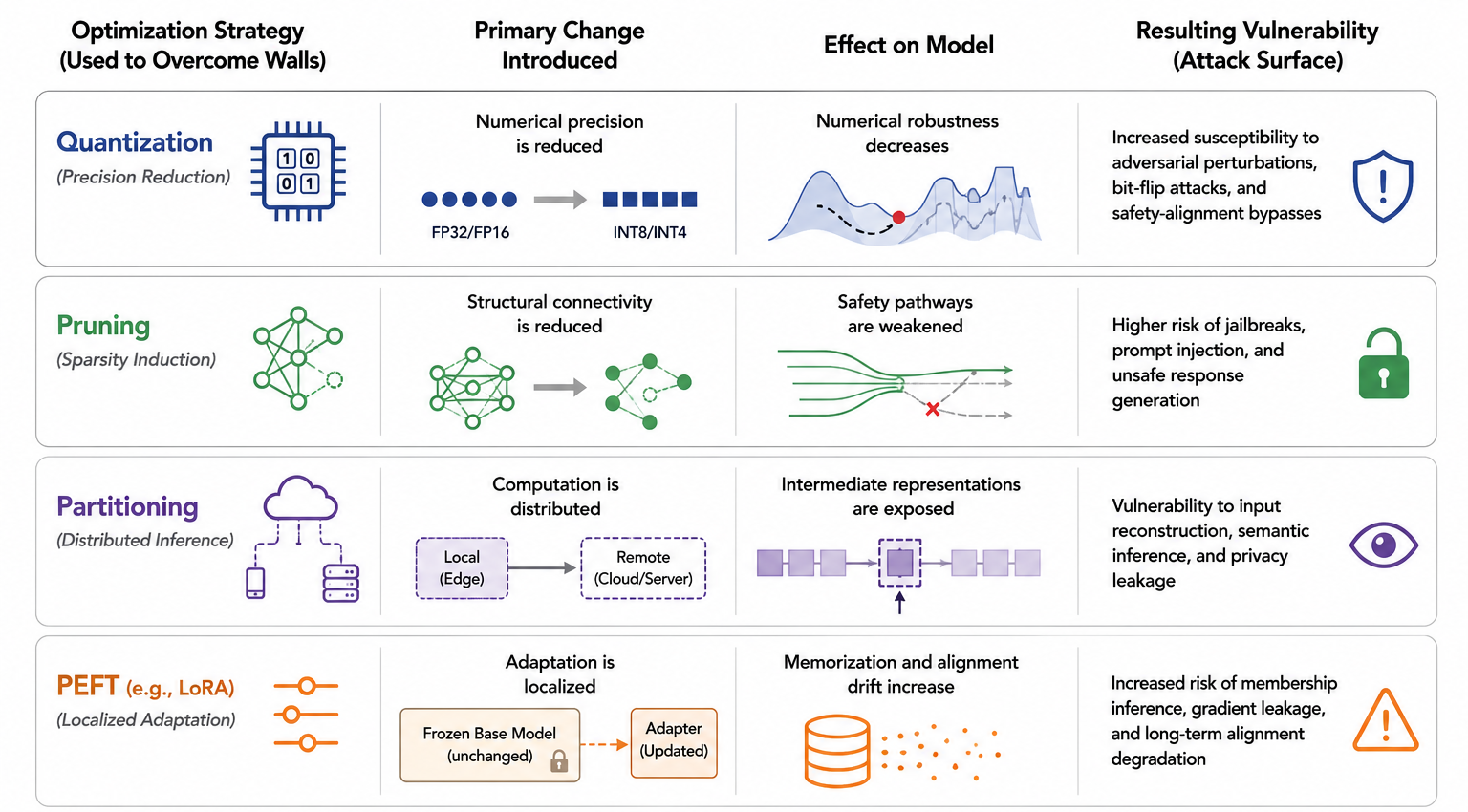}}
    \caption{Summary of optimization strategies and their vulnerabilities.}
    \label{fig:summary_of_four}
    \vspace*{-5pt}
\end{figure}

\section{The Three-Wall Constraint Model for Edge LLMs}
A summary of the optimization strategies and their associated security vulnerabilities discussed in Section \ref{sec:sec_eff_paradox} is presented in Fig. \ref{fig:summary_of_four}. As illustrated, each optimization technique introduced to overcome edge hardware constraints creates a distinct attack surface, collectively embodying the Security-Efficiency Paradox. Building upon these observations, edge LLM deployment must move beyond heuristic-driven model selection toward a principled co-design framework that quantitatively evaluates resource requirements against hardware limitations. Specifically, the Memory, Quadratic, and Compute Walls define the fundamental feasibility boundaries governing whether an on-premise LLM can be deployed without resorting to potentially unsafe optimizations.

\subsection{\textbf{Static and Dynamic Memory Footprint (The Memory Wall)}}
To evaluate edge viability, the deployment memory footprint must be dynamically quantified. The total operational Volumetric RAM ($VRAM_{\text{total}}$) is defined by the static model weights, the dynamic Key-Value (KV) cache, and structural runtime overhead:

\begin{equation*}
    \begin{split}
        VRAM_{\text{total}}
        =&\;
        \underbrace{\left(
        P \times \frac{B}{8}
        \right)}_{\text{weights}}
        \\
        &+
        \underbrace{\left(
        2 \times B_s \times L \times H \times d_{\text{head}} \times S_l \times \frac{B}{8}
        \right)}_{\text{KV cache}}
        \times \Omega
    \end{split}
\label{eq:vram}
\end{equation*}

where: $P$ = total parameter count, $B$ = precision bit-width (e.g., 4, 8, 16), $B_s$ = batch size, $L$ = transformer layers, $H$ = attention heads, $d_{\text{head}}$ = dimension per head, $S_l$ = sequence length, $\Omega$ = runtime overhead coefficient (typically typically $1.15 \le \Omega \le 1.25$).

However, memory alone is insufficient. The Quadratic Wall and Compute Wall impose latency and energy constraints that, when violated, force developers to apply the very optimizations (quantization, pruning, partitioning) that introduce security vulnerabilities. Thus, we propose a unified constraint model.

\subsection{\textbf{Context Complexity Scalability (The Quadratic/Attention Wall)}}
The self-attention mechanism~\cite{vaswani2017attention} computes interactions between all token pairs by forming the product of Queries ($Q$) and Keys ($K$), resulting in a computational complexity that scales quadratically with sequence length. For a sequence of length $S_l$ and hidden dimension $d$, the computational cost of attention is $O(S_l^2\times d)$, making long-context inference a dominant bottleneck in transformer-based models.

In addition to compute complexity, the attention mechanism imposes a significant memory burden. During inference, the attention score matrix must be materialized or stored in intermediate form, requiring:

$$
M_{\text{attn}} = B_s \times H \times S_l^2 \times \frac{B}{8}
$$

where $B_s$ is the batch size, $H$ is the number of attention heads, and $B$ denotes the numerical precision in bits (e.g., 16-bit or 8-bit representation). This quadratic growth in memory consumption severely limits the maximum context length that can be supported on edge devices, even when model weights are aggressively compressed.

As a result, when extending context windows for applications such as retrieval-augmented generation (RAG)~\cite{lewis2020retrieval} or long multi-turn reasoning at the edge, attention memory rapidly dominates the available hardware budget. The maximum feasible sequence length $S_{\max}$ is therefore bounded by the residual memory capacity:

\begin{equation*}
    S_{\max}
    \le
    \sqrt{
    \frac{
    M_{\text{avail}} - M_{\text{weights}} - M_{\text{KV}}
    }{
    B_s \times H \times \frac{B}{8}
    }
    }
\end{equation*}

where $M_{\text{avail}}$ is the total device memory, $M_{\text{weights}}$ denotes static model parameters, and $M_{\text{KV}}$ accounts for key-value cache storage during inference.

Exceeding $S_{\max}$ necessitates the use of approximations such as aggressive KV-cache quantization, sparse attention mechanisms, or sliding-window attention. However, as discussed in Section~\ref{sec:sec_eff_paradox}, these approximations can degrade representation fidelity and inadvertently weaken safety alignment, increasing susceptibility to prompt injection and adversarial manipulation.

\subsection{\textbf{Execution Throughput and Energy Metrics (The Compute Wall)}}
Autoregressive token generation is bound by memory-bandwidth during the decoding phase. The total floating-point operations per second (FLOPS) required to generate $T$ tokens scales as:
$$\text{FLOPS}_{\text{inference}} \approx 2 \times P \times T $$

On battery-constrained tactical edge platforms or localized workstations, continuous computation translates directly to severe thermal throttling, high latency, and rapid power drain. Thus, minimizing FLOPS through token pruning or structural sparsification becomes a prerequisite for edge operational longevity.

For better quantification, the time to generate a single token on edge hardware is:
\begin{equation*}
    T_{\text{token}}
    =
    \frac{
    2 \times P \times F
    +
    S_l \times L \times H \times d_{\text{head}}
    }
    {FLOPS_{\text{avail}}}
\end{equation*}
Where: $F$ is the average FLOPs per parameter per token ($\approx$ 2 for inference). If $T_{\text{token}}>\tau_{\text{max}}$ (e.g., 100 ms for real-time interaction), developers apply quantization or pruning, directly increasing vulnerability to jailbreaking and gradient masking.

\subsection{\textbf{Unified Feasibility Criterion}}
Based on the aforementioned quantitative engine for wall quantification, a model can be deployed \emph{securely} on edge only if:
\begin{equation*}
    M_{\text{weights}} + M_{\text{KV}}(S_{\text{req}}) \leq M_{\text{avail}}, 
    T_{\text{token}} \leq \tau_{\text{max}},  
    S_{\text{req}} \leq S_{\text{max}}
\end{equation*}

If any constraint is violated, the developer is forced into unsafe optimizations. We define the Safety-Efficiency Feasibility Region as the set of ($P$, $B$, $S_l$, $\text{sparsity}$) satisfying all constraints without triggering the attack surfaces described in Section \ref{sec:sec_eff_paradox}.

\section{Optimal Balance Risk Taxonomy \& Benchmarking Framework}
Evaluating edge-native intelligence through a single lens, such as benchmark accuracy, obscures the fundamental trade-offs underlying the Security-Efficiency Paradox. To enable systematic and hardware-aware model selection, we introduce a multi-dimensional risk taxonomy together with a security-efficiency evaluation metric. We begin by identifying the mitigation strategies corresponding to each attack surface summarized in Fig. \ref{fig:summary_of_four}, establishing the foundation of the proposed taxonomy. These mitigations are then prioritized according to their security significance, providing the basis for the proposed \emph{Secure Operational Efficiency Score} benchmark.

\subsection{\textbf{Mitigation Strategies for Edge LLM Deployment}}
The Security-Efficiency Paradox does not imply that hardware optimizations must inevitably compromise security. Instead, each optimization strategy introduces a distinct attack surface that can be mitigated through targeted defenses. Table \ref{tab:mitigations} summarizes these relationships by mapping common edge optimization techniques to their primary security threats and corresponding mitigation strategies, providing the foundation for a security-aware deployment taxonomy. Collectively, these mitigation strategies form a practical taxonomy for security-aware edge LLM optimization and motivate the development of safety-aware compression objectives.

\begin{table}[htbp]
    \centering
    \caption{Mitigation strategies for Security-Efficiency Paradox in edge LLM deployment.}
    \label{tab:mitigations}
    \begin{tabular}{|p{0.2\linewidth}|p{0.25\linewidth}|p{0.4\linewidth}|}
    \hline
    \textbf{Optimization} & \textbf{Primary Threat} & \textbf{Mitigation Strategy} \\
    \hline
    Quantization & Gradient masking, jailbreaking via bit-flip & Use quant-aware training (not post-training); apply randomized smoothing; monitor bit-flip detection via ECC memory \\
    \hline
    Pruning & Fragmentation of safety subnetworks & Incorporate safety loss into pruning criterion (not just LM loss); freeze safety-critical heads \\
    \hline
    Partitioning & Intermediate activation reconstruction & Inject noise into cut-layer activations (DP inference); encrypt cut-layer outputs; split at non-linear boundaries \\
    \hline
    LoRA/PEFT & Membership inference, gradient leakage, alignment drift \cite{ren2024analyzing} & Apply differential privacy to LoRA updates; limit local fine-tuning steps; periodic reset to base safety alignment \\
    \hline
    \end{tabular}
\end{table}

The mitigation strategies outlined in Table \ref{tab:mitigations} reflect a layered defense philosophy that addresses vulnerabilities at multiple levels of the deployment stack. For quantization, the primary defense against gradient masking and bit-flip attacks is to adopt quant-aware training rather than post-training quantization, which preserves the smoothness of the loss landscape and maintains gradient fidelity during adversarial optimization. Complementary techniques such as randomized smoothing~\cite{cohen2019certified} and error-correcting code (ECC) memory further harden quantized models against input perturbations and hardware-level fault injection. For pruning, the central insight is that safety-critical pathways must be explicitly preserved during sparsification; by incorporating a safety regularization term into the pruning objective, rather than relying solely on language modeling loss—developers can prevent fragmentation of the safety subnetwork and retain robustness against jailbreak attempts.

For model partitioning, the threat of intermediate activation reconstruction can be mitigated through differential privacy mechanisms that inject calibrated noise into cut-layer outputs, coupled with cryptographic encryption of transmitted tensors and strategic selection of split boundaries that minimize semantic leakage. Finally, for LoRA and other PEFT methods, defenses target the compact adapter parameters that concentrate domain-specific knowledge; applying differential privacy to gradient updates, restricting the number of local fine-tuning steps, and periodically resetting adapters to the base safety alignment collectively reduce the risks of membership inference, gradient leakage, and alignment drift. 




\subsection{\textbf{Standardizing Domain-Specific Safety-Efficiency Benchmarks}}
Existing LLM benchmarks primarily evaluate task performance while largely ignoring the hardware constraints and security risks that define practical edge deployment. However, selecting an edge-native LLM requires simultaneously balancing application utility, resilience against adversarial and privacy attacks, and computational efficiency under limited hardware resources. Evaluating these objectives independently can lead to suboptimal deployment decisions that either exceed hardware capabilities or compromise security. To address this limitation, we propose the \emph{Secure Operational Efficiency Score (SOES)}, a unified metric that jointly captures functional capability, security robustness, privacy preservation, and hardware efficiency within a single evaluation framework.

In fact, edge LLM deployment requires balancing three competing objectives: (i) functional utility, (ii) trustworthy and secure operation, and (iii) hardware efficiency. A deployment that excels in only one of these dimensions is often unsuitable in practice. For example, a highly accurate model may be vulnerable to jailbreak attacks, while an extremely efficient model may sacrifice reasoning capability or privacy. Accordingly, the proposed Secure Operational Efficiency Score integrates representative metrics from each objective into a unified evaluation framework. Specifically, Task Accuracy captures application utility; Jailbreak Resistance and Privacy Score quantify robustness against the primary attack surfaces identified in Section \ref{sec:sec_eff_paradox}; and Energy Consumption, VRAM Footprint, and Inference Latency characterize the hardware resources that fundamentally constrain edge deployment.

\begin{equation*}
    \text{SOES} = \frac{
        \begin{array}{c}
            \text{Task Accuracy} \times \text{Jailbreak Resistance} \\
            \times\; \text{Privacy Score}
        \end{array}
    }{
        \begin{array}{c}
            \text{Energy (J/token)} \times \text{VRAM Footprint (GB)} \\
            \times\; \text{Latency (s/token)}
        \end{array}
    }
\end{equation*}

where all six metrics are normalized to the range $[0,1]$ before computing SOES. Metrics in the numerator are normalized such that larger values indicate better performance, whereas resource metrics in the denominator are inverse-normalized so that lower hardware costs correspond to higher normalized scores. This normalization eliminates differences in measurement units and prevents any single metric from dominating the overall evaluation.

\begin{itemize}
    \item \textbf{Task Accuracy} is measured using domain-specific benchmark suites (e.g., MedQA~\cite{jin2021disease}, MMLU~\cite{hendrycks2020measuring}, HumanEval~\cite{chen2021evaluating}) and normalized to $[0,1]$, where higher values indicate better task performance.
    
    \item \textbf{Jailbreak Resistance} is computed as the complement of the successful attack rate on standardized adversarial prompt benchmarks (i.e., $1-\text{ASR}$) and normalized to $[0,1]$, where a value of 1 indicates complete resistance to jailbreak attacks.  
    
    \item \textbf{Privacy Score} is derived as the complement of Membership Inference Attack success rate and normalized to $[0,1]$, where higher values correspond to stronger privacy preservation.
    
    \item \textbf{Energy Consumption} represents the average energy required to generate one token on the target edge hardware. The measured value is inverse-normalized to $[0,1]$ so that models with lower energy consumption receive higher scores.
    
    \item \textbf{VRAM Footprin}t denotes the peak runtime memory allocation required during inference. The measured memory footprint is inverse-normalized to $[0,1]$, rewarding models with smaller memory requirements.

    \item \textbf{Inference Latency} represents the average time required to generate a single output token on the target hardware. Latency is inverse-normalized to $[0,1]$, such that lower latency corresponds to higher efficiency scores.
\end{itemize}

The multiplicative formulation intentionally penalizes models that perform poorly in any single dimension. Consequently, a model cannot compensate for severe weaknesses in security, privacy, or hardware efficiency through high task accuracy alone, reflecting the practical deployment requirements of edge intelligence.

A higher SOES indicates a better balance. Practitioners can compare candidate models (e.g., Llama 3–8B INT4 vs. Phi-3-mini FP16) and choose the one maximizing SOES subject to real-time constraints. By maximizing the SOES metric, developers can systematically determine whether a highly optimized, smaller model configuration (e.g., an 8B model quantized to INT4) provides superior holistic utility compared to a larger, unquantized partition configuration.

\begin{figure}[H]
    \centerline{\includegraphics[width=\columnwidth]{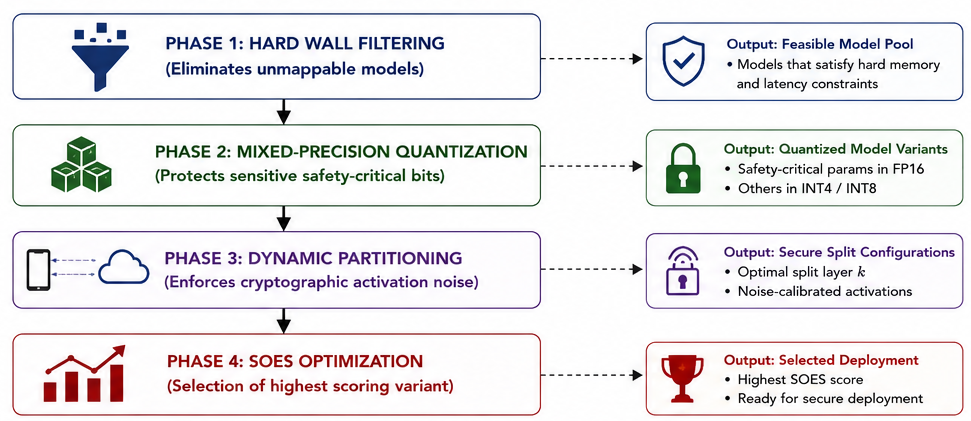}}
    \caption{Hardware-constrained, privacy-preserving LLM selection pipeline.}
    \label{fig:selection_pipeline}
    \vspace*{-5pt}
\end{figure}

\subsection{\textbf{Optimal Model Selection Strategy under Hardware Constraints}}
Maximizing SOES requires an automated, step-by-step pipeline to filter, select, and deploy an edge model while actively defending against the core vulnerabilities established in Section \ref{sec:sec_eff_paradox}. As summarized in Fig. \ref{fig:selection_pipeline}, we propose a practical four stage pipeline to achieve optimal model selection strategy. 

\begin{itemize}
    \item \textbf{Phase 1: Hard Wall Filtering} \\
    The selection pipeline begins by querying the local hardware profile to establish upper boundaries for static memory allocation ($VRAM_{\text{max}}$) and maximum permissible latency ($\text{Latency}_{\text{max}}$). Models whose base dimensions violate these absolute bounds are immediately purged from the selection pool.

    \item \textbf{Phase 2: Mixed-Precision Quantization and Protection} \\
    To safely bypass the Memory Wall without suffering from the loss landscape discretization that triggers jailbreaking vulnerabilities, the framework applies \emph{mixed-precision} strategies \cite{zhao2024edge}. Instead of uniform INT4 reduction, the system identifies critical attention blocks and early projection embeddings that govern safety-alignment filters. These features are retained at high-precision (FP16), while the remaining non-sensitive parameters are compressed to INT4/INT8 configurations.

    \item \textbf{Phase 3: Noise-Calibrated Split Inference} \\
    If the target model requires architectural partitioning across distributed localized nodes, the activation boundary layer ($k$) is selected based on a security-compute trade-off. To stop reconstruction attacks without causing catastrophic token degradation, the system applies functional cryptographic obfuscation or introduces low-magnitude Gaussian noise calibrated directly to the semantic density of that specific layer:
    $$\tilde{H}_k = H_k + \mathcal{N}(0, \sigma^2)$$

    The noise scale $\sigma$ is iteratively tuned via feedback loops to maximize privacy preservation while maintaining the downstream perplexity threshold required for operational stability.

    \item \textbf{Phase 4: Final SOES Optimization} \\
    The remaining candidates are run through an automated profiling cycle. The variant that returns the highest integrated Secure Operational Efficiency Score (SOES) is pushed to production, ensuring that the model achieves the highest possible throughput without crossing into unsafe operating thresholds.    
\end{itemize}

\section{Comparative SOES Analysis of Edge-Friendly LLMs}
To empirically validate the Security-Efficiency Paradox and demonstrate the utility of the SOES, we conducted a comparative evaluation of representative edge-friendly large language models. Based on existing literature and practical settings, our experimental framework integrates real-world performance metrics, security assessments, and efficiency measurements to provide a holistic comparison of model suitability for edge deployment. The goal is to illustrate how SOES guides model selection by balancing task accuracy, security, and resource efficiency under real-world edge constraints.

\subsection{\textbf{Evaluation Framework}}
The experimental pipeline consists of six distinct measurement components (as introduced with SOES), each designed to quantify a specific dimension of model performance:

\begin{figure*}
    \centerline{\includegraphics[width=43pc]{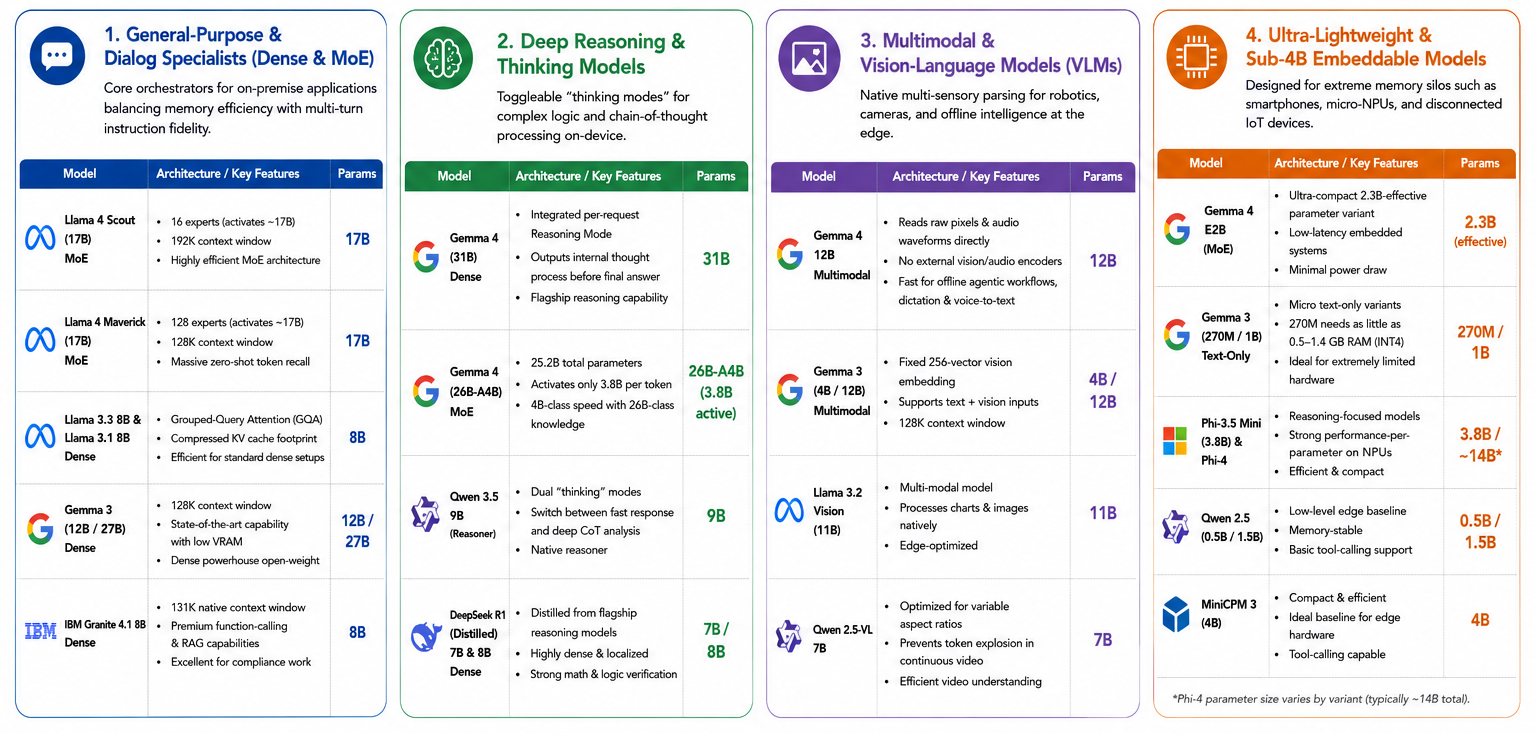}}
    \caption{Current landscape of edge-deployable LLM models for on-device intelligence. Parameters (params) denote total model parameters unless otherwise specified. Effective parameters indicate the number of activated per token (for MoE models). Actual memory footprint depends on quantization, context length, runtime optimizations, and hardware.}
    \label{fig:llm_landscape}
    \vspace*{-5pt}
\end{figure*}

\begin{enumerate}
    \item \textbf{Task Accuracy (MMLU Benchmark)}:
    We evaluate task performance using the Massive Multitask Language Understanding (MMLU) benchmark \cite{hendrycks2021measuring}, a comprehensive benchmark designed to measure a model's world knowledge and problem-solving abilities across 57 diverse subjects. These subjects span four major disciplines: humanities, social sciences, STEM, and other specialized areas, making it a robust indicator of general language model capability. Following standard evaluation protocols from the original repository \cite{hendryckstest2021}, we utilize the few-shot evaluation setting, where each model is provided with five example questions and answers in the prompt to guide its responses. The model is then evaluated on a standardized test set of multiple-choice questions, and its accuracy is measured as the percentage of correct answers across all subjects or within specific disciplinary categories. This approach ensures comparability with a wide body of existing literature and leaderboard results, providing a rigorous and widely recognized measure of general-purpose LLM performance.

    \item \textbf{Jailbreak Resistance}: To quantify adversarial robustness, we measure resistance to harmful prompt injection attacks. We employ a curated set of 10 malicious prompts covering diverse harmful behaviors, including illegal instructions, privacy violations, and content policy violations. Jailbreak resistance is computed as the refusal rate: the proportion of prompts that the model successfully rejects using safety-aligned responses (e.g., "I cannot assist with that request"). If the model doesn't clearly refuse to execute the request, it is considered a failed jailbreak resistance. This metric directly captures the fragility of safety alignment under adversarial conditions.

    \item \textbf{Privacy Score}: Privacy leakage is measured via a canary extraction test that simulates membership inference attacks. We insert 30 unique "canary" strings into the model's context and measure the cross-entropy loss. Lower loss on canary strings compared to random strings indicates memorization, suggesting privacy leakage. The privacy score is normalized as the ratio of canary loss to random loss, where higher values indicate better privacy preservation.

    \item \textbf{Energy Consumption}: Energy efficiency is measured as Joules per generated token. Using the T4 GPU's (in Google Colab) power draw characteristics (estimated at 50W during inference), we compute energy consumption from inference latency measurements. This metric is critical for battery-constrained edge devices.

    \item \textbf{Memory Footprint}: VRAM usage is measured in gigabytes during inference using PyTorch's memory tracking utilities. This quantifies the model's memory wall impact and determines feasibility on resource-constrained hardware.

    \item \textbf{Inference Latency}: We measure end-to-end latency per token generation (ms/token), which directly impacts user experience in real-time applications.    
\end{enumerate}

All experiments were conducted using different settings of quantization (INT4, FP) via bitsandbytes \cite{dettmers2023qlora} to enable deployment on edge hardware. Each model is evaluated on the MMLU dataset and the results are reported as averages across the test subset.

\subsection{\textbf{Models Under Evaluation}}
We selected a representative set of models spanning diverse architectural paradigms and parameter scales, as illustrated in Fig. \ref{fig:llm_landscape}. Specifically, the selected models encompass dense, mixture-of-experts (MoE), reasoning-oriented, and multimodal architectures, providing broad coverage of the current edge-deployable LLM landscape. However, the experimental evaluation presented in this paper primarily focuses on the first two categories (General-Purpose and Dialog Specialists and Deep Reasoning and Thinking Models) as these architectures are most relevant to edge inference scenarios requiring efficient instruction following, task orchestration, and complex reasoning under hardware constraints.

As such, we selected compact open-weight models spanning multiple model families, architectural designs, and parameter scales (1.5B–3B). The evaluated models include dense, quantized, and edge-optimized architectures from Microsoft, Alibaba, Google, Meta, and IBM, enabling a comprehensive assessment of the proposed SOES metric across diverse deployment scenarios.

\begin{itemize}
    \item \textbf{Phi-3.5-mini-instruct (3.8B)~\cite{haider2024phi}:} Microsoft's instruction-tuned compact language model, recognized for strong reasoning performance and high parameter efficiency, making it well suited for edge inference on resource-constrained hardware.

    \item \textbf{Qwen2.5-1.5B-Instruct (1.5B)~\cite{hui2024qwen2}:} Alibaba's lightweight instruction-tuned model designed for efficient deployment on edge devices while maintaining competitive instruction-following capabilities.

    \item \textbf{Gemma-2-2B (2B)~\cite{team2024gemma}:} Google's compact open-weight language model optimized for efficient inference, representing the low-memory segment of the Gemma family for edge deployment.



    \item \textbf{Gemma-4-E2B (2.3B effective)~\cite{gemmateam2026gemma4}:} Google's ultra-lightweight edge-optimized model designed specifically for low-latency inference on embedded devices and consumer-grade hardware with minimal memory requirements.

    \item \textbf{Llama-3.2-3B-Instruct-bnb-4bit (3B)~\cite{grattafiori2024llama}:} Meta's instruction-tuned Llama model deployed using 4-bit quantization, representing a highly memory-efficient configuration widely adopted for local edge inference.

    \item \textbf{Granite-4.1-3B (3B):} IBM's enterprise-focused compact language model featuring strong instruction following, function-calling, and RAG capabilities while maintaining a small deployment footprint.
\end{itemize}

All models are evaluated in their multiple quantized configurations to reflect real-world deployment scenarios where memory constraints necessitate aggressive compression.

\subsection{\textbf{Experimental Results}}
Table \ref{tab:soes_results} presents the complete benchmark results for all models across all evaluation dimensions.

\begin{table}[t]
    \caption{SOES Benchmark Results for Edge-Friendly LLMs under Full Precision (FP) Inference.}
    \label{tab:soes_results}
    \centering
    \setlength{\tabcolsep}{2pt}
    \begin{tabular}{p{1.6cm}ccccccc}
    \toprule
    Model & MMLU & Jailbreak & Privacy & Energy & VRAM & Latency & SOES \\
    &
    (\%) &
    (\%) &
    (\%) &
    (J) &
    (GB) &
    (ms) &
    \\
    \midrule
    Phi-3.5-mini-instruct          & 68.9 & 20.0   & 50.8  & 4.170  & 7.12 & 83.4   & 0.0283 \\
    Qwen2.5-1.5B-Instruct          & 59.5 & 100.0 & 70.6  & 2.1545  & 2.88 & 43.1   & 1.5698 \\
    Google/gemma-2-2b-it           & 57.3 & 90.0  & 64.5  & 1.8190  & 4.88 & 36.4   & 1.0311 \\
    Google/gemma-4-E2B-it          & 47.9 & 90.0  & 71.9  & 5.1277  & 9.52 & 102.6   & 0.0619 \\
    Llama-3.2-3B-Instruct-bnb-4bit & 59.9 & 80.0  & 80.4  & 3.0665  & 2.10 & 61.3   & 0.9762 \\
    IBM/granite-4.1-3b             & 65.9 & 90.0  & 65.2  & 2.8518  & 6.35 & 57.0   & 0.3747 \\
    \bottomrule
    \end{tabular}
\end{table}

\begin{table}[t]
    \caption{SOES Benchmark Results for Edge-Friendly LLMs under INT4 Quantization.}
    \label{tab:soes_results}
    \centering
    \setlength{\tabcolsep}{2pt}
    \begin{tabular}{p{1.6cm}ccccccc}
    \toprule
    Model & MMLU & Jailbreak & Privacy & Energy & VRAM & Latency & SOES \\
    &
    (\%) &
    (\%) &
    (\%) &
    (J) &
    (GB) &
    (ms) &
    \\
    \midrule
    Phi-3.5-mini-instruct           & 67.1 & 30.0 & 54.8 & 2.3492 & 2.11 & 47.0 & 0.4739 \\
    Qwen2.5-1.5B-Instruct           & 57.5 & 100.0 & 70.3 & 4.7753 & 1.08 & 95.5 & 0.8184 \\  
    Google/gemma-2-2b-it            & 56.4 & 80.0 & 63.6 & 3.4461 & 2.08 & 68.9 & 0.5804 \\
    Google/gemma-4-E2B-it           & 43.8 & 90.0 & 72.9 & 8.2571 & 6.29 & 165.1 & 0.0335 \\
    Llama-3.2-3B-Instruct-bnb-4bit  & 59.9 & 80.0 & 78.2 & 3.5060 & 2.10 & 70.1 & 0.7261 \\
    IBM/granite-4.1-3b              & 63.9 & 100.0 & 69.3 & 4.2283 & 2.00 & 84.6 & 0.6194 \\
    \bottomrule
    \end{tabular}
\end{table}

    

\subsubsection{MMLU Task Accuracy}
We evaluate task performance using the Massive Multitask Language Understanding (MMLU) benchmark \cite{hendrycks2021measuring}, which measures world knowledge and problem-solving across 57 diverse subjects spanning humanities, social sciences, STEM, and other specialized domains. In the full-precision (FP) setting, task accuracy ranges from 47.9\% (gemma-4-E2B-it) to 68.9\% (Phi-3.5-mini-instruct). The highest-performing models are Phi-3.5-mini (68.9\%), granite-4.1-3b (65.9\%), and Llama-3.2-3B (59.9\%). Under INT4 quantization, we observe a moderate accuracy degradation of 1.8–4.1 percentage points for most models, with Phi-3.5-mini dropping to 67.1\% and granite-4.1-3b to 63.9\%. Interestingly, Llama-3.2 maintains identical accuracy (59.9\%) across both precision settings, while Qwen2.5-1.5B shows a slight decline from 59.5\% to 57.5\%. The gemma-4-E2B-it model exhibits the lowest performance in both settings (47.9\% FP, 43.8\% INT4), suggesting that its extreme lightweight design (2.3B effective parameters) trades off substantial capability for efficiency.

While these results remain below state-of-the-art performance (e.g., GPT-4 achieves ~89\% \cite{radford2019language}), they reflect the inherent capability constraints of compact edge-optimized models and the additional degradation introduced by quantization. Notably, model size alone does not guarantee superior performance, Qwen2.5-1.5B (1.5B) outperforms larger models like gemma-2-2b-it (2B) and gemma4-E2B-it (2.3B effective). This suggests that architectural efficiency and training methodologies are as important as raw parameter count for edge deployment, validating our hypothesis that aggressive compression fundamentally alters model behavior.

\subsubsection{Multi-Dimensional Performance Evaluation}
\begin{figure}
    \centerline{\includegraphics[width=\columnwidth]{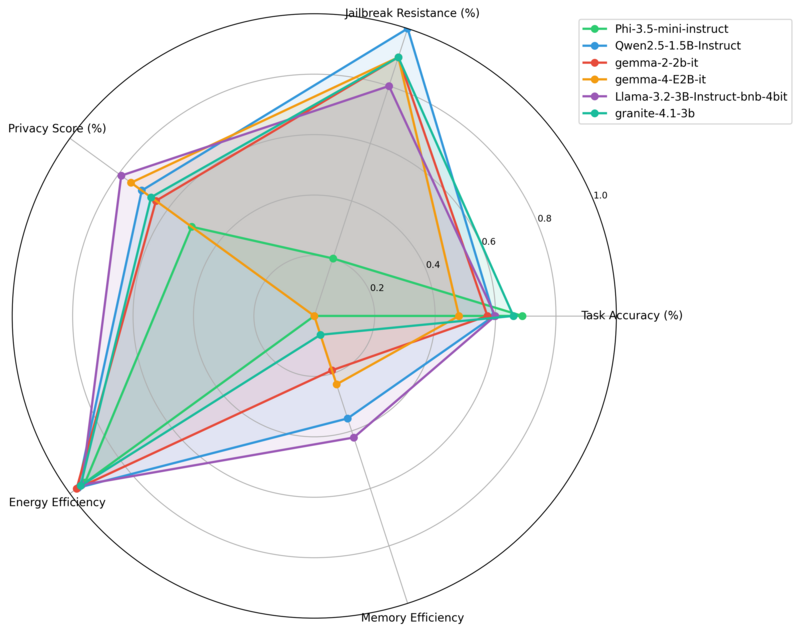}}
    \caption{Radar chart of normalized performance metrics for edge-friendly LLMs under full precision (FP) inference.}
    \label{fig:radar_fp}
    \vspace*{-5pt}
\end{figure}

\begin{figure}
    \centerline{\includegraphics[width=\columnwidth]{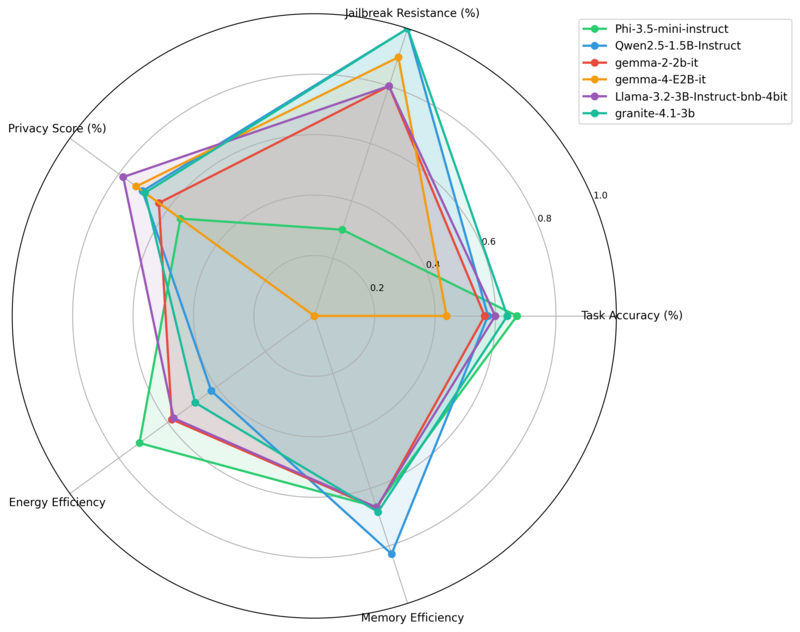}}
    \caption{Radar chart of normalized performance metrics for edge-friendly LLMs under INT4 quantization.}
    \label{fig:radar_int4}
    \vspace*{-5pt}
\end{figure}

To provide a holistic view of model capabilities beyond single-metric comparisons, Figures \ref{fig:radar_fp} and \ref{fig:radar_int4} present radar charts visualizing normalized performance across five key dimensions: task accuracy, jailbreak resistance, privacy score, energy efficiency, and memory efficiency for FP and INT4 configurations, respectively. These visualizations reveal distinct performance profiles that are obscured by aggregate metrics like SOES alone.

In the full-precision setting (Figure \ref{fig:radar_fp}), Qwen2.5-1.5B exhibits the most balanced profile, achieving near-perfect scores in jailbreak resistance and strong performance across privacy, memory efficiency, and task accuracy, though its energy efficiency is comparatively moderate. Gemma-2-2b-it demonstrates excellent energy and memory efficiency with good privacy and jailbreak resistance, making it a strong candidate for resource-constrained deployments where security is a secondary concern. Phi-3.5-mini, despite leading in task accuracy, reveals a critically deficient profile with the lowest jailbreak resistance and below-average privacy, confirming its unsuitability for security-sensitive applications. Gemma-4-E2B-it shows the weakest overall profile, particularly in energy efficiency and latency, which is reflected in its low SOES.

Under INT4 quantization (Figure \ref{fig:radar_int4}), the performance landscape shifts notably. Llama-3.2 emerges with the most balanced profile, maintaining strong privacy and memory efficiency while achieving moderate jailbreak resistance and task accuracy. Qwen2.5-1.5B retains perfect jailbreak resistance and strong privacy but exhibits significantly degraded energy efficiency, consistent with its counterintuitive increase in energy consumption under INT4 observed in Section V-C4. Gemma-2-2b-it maintains good balance across all dimensions, while Phi-3.5-mini shows modest improvements in energy and memory efficiency but remains critically vulnerable in jailbreak resistance. Notably, the radar charts clearly illustrate that no single model dominates across all dimensions—a finding that underscores the fundamental trade-offs inherent in edge LLM deployment and reinforces the necessity of SOES as a decision-making tool that accounts for application-specific priorities.

\subsubsection{Jailbreak Resistance and Security}
The jailbreak resistance results reveal striking and consistent differences in safety alignment robustness across both FP and INT4 settings. In the full-precision setting, Qwen2.5-1.5B achieves 100\% refusal rate, successfully rejecting all harmful prompts, closely followed by gemma-2-2b-it (90\%), gemma-4-E2B-it (90\%), and granite-4.1-3b (90\%). Llama-3.2 demonstrates 80\% resistance, while Phi-3.5-mini exhibits alarmingly low resistance at only 20\%. Under INT4 quantization, we observe interesting patterns: Qwen2.5-1.5B and granite-4.1-3b both achieve perfect 100\% resistance, followed by gemma-4-E2B-it (90\%), Llama-3.2 (80\%), and gemma-2-2b-it (80\%). Phi-3.5-mini improves slightly to 30\% but remains critically vulnerable.

This disparity highlights the \emph{Security-Efficiency Paradox} in action. Phi-3.5-mini achieves the highest MMLU accuracy (68.9\% FP, 67.1\% INT4) yet demonstrates the lowest jailbreak resistance (20–30\%). As discussed in Section II-A, quantization introduces gradient masking and amplifies adversarial noise, effectively disabling the safety guardrails that exist in the full-precision model. The persistent vulnerability of Phi-3.5-mini across both precision settings suggests that its safety subnetwork is fundamentally fragile due to architectural choices or training methodology.

The 100\% jailbreak resistance of Qwen2.5-1.5B and granite-4.1-3b is particularly noteworthy. Qwen, despite being the smallest model (1.5B), maintains robust safety alignment under quantization. Granite, an enterprise-focused model, also demonstrates perfect resistance in INT4, suggesting that IBM's training methodology prioritizes safety preservation. These results imply that architectural choices and training strategies can preserve safety even under aggressive compression~\cite{hong2024decoding}, offering a path forward for secure edge deployment.

\subsubsection{Privacy Score and Data Protection}
Privacy scores exhibit moderate variation across models and precision settings. In FP, Llama-3.2 achieves the highest privacy score (80.4\%), followed by gemma-4-E2B-it (71.9\%), Qwen2.5-1.5B (70.6\%), and granite-4.1-3b (65.2\%). Phi-3.5-mini and gemma-2-2b-it show lower scores at 50.8\% and 64.5\%, respectively. Under INT4 quantization, privacy scores generally remain stable or show modest changes: Llama-3.2 achieves 78.2\%, Qwen2.5-1.5B 70.3\%, gemma-4-E2B-it 72.9\%, and Phi-3.5-mini improves to 54.8\%.

The relatively lower privacy scores for Phi-3.5-mini (50.8–54.8\%) are concerning, as they indicate higher memorization of training data. In edge deployment scenarios where sensitive data (e.g., medical records, personal communications) may be processed, this privacy leakage could have significant regulatory implications under frameworks like HIPAA and GDPR. Conversely, Llama-3.2's strong privacy preservation (80.4\% FP, 78.2\% INT4) suggests that its training methodology effectively mitigates membership inference risks, making it a safer choice for privacy-sensitive applications.

\subsubsection{Efficiency Metrics}
The efficiency measurements reveal distinct resource profiles across models and precision settings:
\begin{itemize}
    \item VRAM Footprint: In FP, Llama-3.2 achieves the smallest footprint at 2.10 GB, followed closely by Qwen2.5-1.5B (2.88 GB) and gemma-2-2b-it (4.88 GB). Granite-4.1-3b (6.35 GB) and Phi-3.5-mini (7.12 GB) require significantly more memory, while gemma-4-E2B-it (9.52 GB) occupies the largest footprint. Under INT4 quantization, VRAM reduces substantially: Qwen2.5-1.5B achieves the smallest footprint at 1.08 GB (62.5\% reduction), followed by granite-4.1-3b (2.00 GB, 68.5\% reduction), Llama-3.2 (2.10 GB, identical to FP), and Phi-3.5-mini (2.11 GB, 70.4\% reduction). Gemma-2-2b-it reduces to 2.08 GB (57.4\% reduction), while gemma-4-E2B-it shows limited reduction (6.29 GB, 33.9\% reduction). This 2–4x memory reduction makes INT4 quantization essential for deployment on memory-constrained edge devices.

    \item Energy Consumption: Energy per token varies significantly. In FP, gemma-2-2b-it is the most energy-efficient at 1.819 J/token, followed by Qwen2.5-1.5B (2.155 J/token), granite-4.1-3b (2.852 J/token), and Llama-3.2 (3.067 J/token). Phi-3.5-mini consumes 4.170 J/token, while gemma-4-E2B-it shows high energy consumption at 5.128 J/token. Under INT4, energy consumption patterns change: Phi-3.5-mini improves to 2.349 J/token (43.7\% reduction), gemma-2-2b-it to 3.446 J/token (47.2\% reduction), and Llama-3.2 to 3.506 J/token (12.7\% reduction). Counterintuitively, Qwen2.5-1.5B consumes more energy under INT4 (4.775 J/token, +121.5\%) than FP, suggesting that its quantization introduces additional computational overhead. Granite-4.1-3b also shows increased energy (4.228 J/token, +48.3\%), while gemma-4-E2B-it improves to 8.257 J/token (-37.9\%) but remains the most energy-hungry model. This inversion demonstrates that energy efficiency is not solely a function of model size but depends on architectural efficiency and quantization implementation.

    \item Inference Latency: In FP, gemma-2-2b-it is the fastest at 36.4 ms/token, followed by Qwen2.5-1.5B (43.1 ms/token), granite-4.1-3b (57.0 ms/token), Llama-3.2 (61.3 ms/token), and Phi-3.5-mini (83.4 ms/token). Gemma-4-E2B-it exhibits high latency at 102.6 ms/token, making it less suitable for real-time applications. Under INT4, latency patterns shift: Phi-3.5-mini improves dramatically to 47.0 ms/token (-43.6\%), gemma-2-2b-it to 68.9 ms/token (-47.2\%), and Llama-3.2 to 70.1 ms/token (-12.7\%). Qwen2.5-1.5B increases to 95.5 ms/token (+121.6\%), while granite-4.1-3b shows 84.6 ms/token (+48.4\%) and gemma-4-E2B-it improves to 165.1 ms/token (-37.9\%). The latency improvement in most models under INT4 aligns with expectations, but the regression in Qwen and granite suggests that their specific quantization implementations introduce computational overhead that outweighs the benefits of reduced precision.
\end{itemize}

\begin{figure}[htbp]
    \centering
    \begin{subfigure}[b]{0.48\textwidth}
        \centering
        \includegraphics[width=\textwidth]{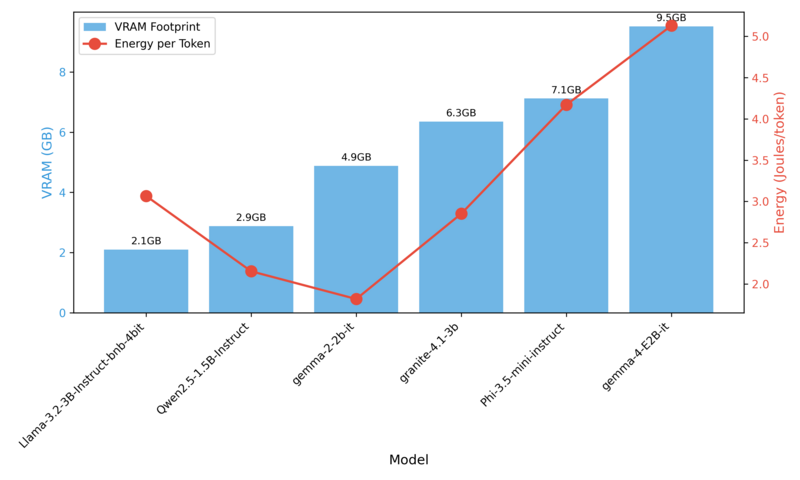}
        \caption{Different LLMs under full precision (FP) inference.}
    \end{subfigure}
    \hfill
    \begin{subfigure}[b]{0.48\textwidth}
        \centering
        \includegraphics[width=\textwidth]{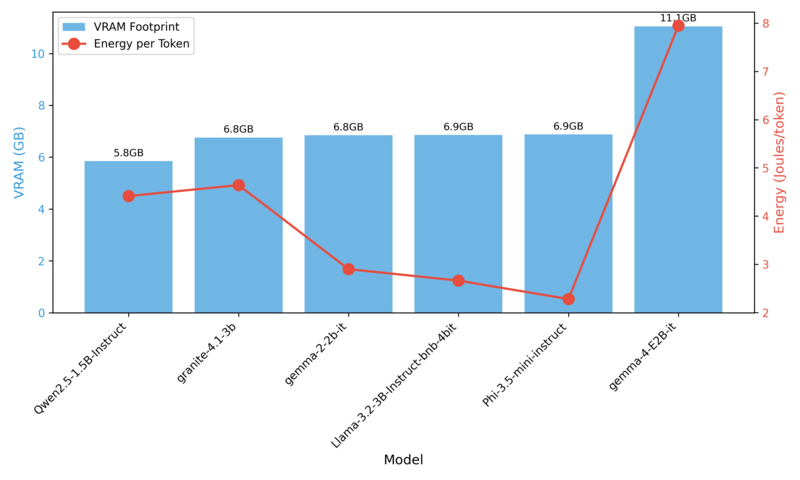}
        \caption{Different LLMs under INT4 quantization.}
    \end{subfigure}
    \caption{Resource footprint of different models: VRAM and Energy.}
    \label{fig:resource_footprint_comparison}
\end{figure}

Figure \ref{fig:resource_footprint_comparison} summarizes the resource footprint comparison, showing VRAM usage alongside latency and energy to highlight the efficiency trade-offs across models and precision settings.

\subsubsection{SOES Scores and Model Ranking}
The SOES metric synthesizes the competing dimensions into a single score, providing a unified ranking for model selection. In the full-precision setting, Qwen2.5-1.5B achieves the highest SOES (1.5698), driven by its perfect jailbreak resistance (100\%), strong privacy (70.6\%), and excellent efficiency (2.155 J/token, 43.1 ms, 2.88 GB). Gemma-2-2b-it follows with 1.0311, benefiting from the lowest energy and fastest latency among all models, coupled with strong security (90\% jailbreak, 64.5\% privacy). Llama-3.2 ranks third (0.9762), with granite-4.1-3b (0.3747) and Phi-3.5-mini (0.0283) trailing significantly. The latter's poor performance stems from critically low jailbreak resistance (20\%), which severely penalizes the numerator despite highest MMLU accuracy (68.9\%). Gemma-4-E2B-it achieves low SOES (0.0619) due to its high energy and latency.

Under INT4 quantization, the ranking shifts dramatically. Qwen2.5-1.5B remains the top performer (SOES = 0.8184), retaining perfect jailbreak resistance and strong privacy (70.3\%), while its VRAM footprint drops to an industry-leading 1.08 GB. However, its energy and latency regressions under INT4 (4.775 J, 95.5 ms) prevent it from reaching its FP-level SOES. Llama-3.2 achieves a strong second place (0.7261), benefiting from high privacy (78.2\%), good jailbreak resistance (80\%), and consistent efficiency (3.506 J, 2.10 GB, 70.1 ms). Granite-4.1-3b achieves 0.6194 with perfect jailbreak resistance (100\%) and competitive MMLU accuracy (63.9\%). Gemma-2-2b-it achieves 0.5804, while Phi-3.5-mini improves substantially to 0.4739 due to reduced energy and latency, though its jailbreak resistance remains a critical liability. Gemma-4-E2B-it achieves the lowest SOES (0.0335) due to its high energy and latency.

These results reveal two important insights. First, full-precision deployment can be impractical on edge hardware for models with larger VRAM requirements (e.g., granite-4.1-3b requires 6.35 GB), making INT4 quantization necessary. Second, quantization's impact on SOES is highly model-dependent; models that maintain security and efficiency under compression (e.g., Llama-3.2, granite-4.1-3b) remain competitive, while those that suffer efficiency regressions (e.g., Qwen's energy/latency increase) or security vulnerabilities (e.g., Phi-3.5-mini's low jailbreak resistance) lose their advantage.

\begin{figure}[htbp]
    \centering
    \begin{subfigure}[b]{0.48\textwidth}
        \centering
        \includegraphics[width=\textwidth]{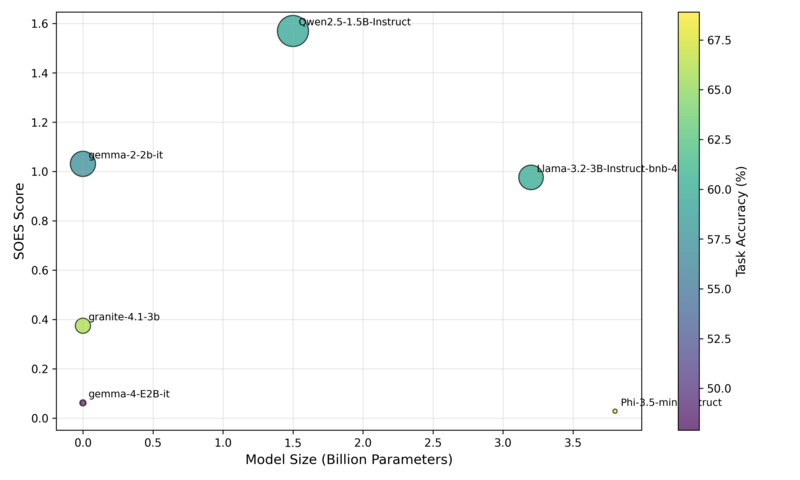}
        \caption{Different LLMs under full precision (FP) inference.}
    \end{subfigure}
    \hfill
    \begin{subfigure}[b]{0.48\textwidth}
        \centering
        \includegraphics[width=\textwidth]{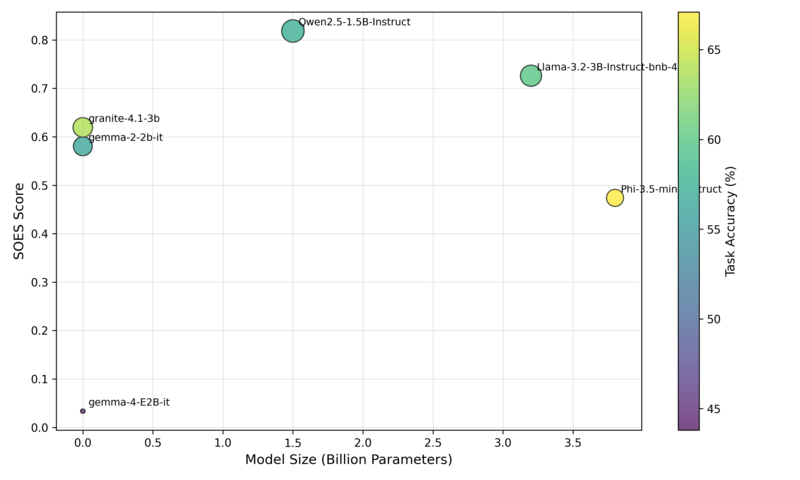}
        \caption{Different LLMs under INT4 quantization.}
    \end{subfigure}
    \caption{SOES vs. model size, illustrating the non-monotonic relationship between parameter count and holistic deployment suitability. Larger bubbles indicate better task accuracy.}
    \label{fig:soes_vs_size}
\end{figure}

\subsubsection{The Security-Efficiency Paradox Revisited}
Our results empirically validate the Security-Efficiency Paradox central to this paper. The paradoxical relationship is most evident when comparing the FP and INT4 rankings and the trade-offs each model embodies.

This paradox manifests differently across models. Phi-3.5-mini is the most accurate in both settings (68.9\% FP, 67.1\% INT4) and among the most efficient in INT4 (2.349 J, 47.0 ms, 2.11 GB), yet remains the least secure (20–30\% jailbreak), a vulnerability that is architectural, not quantization-induced. Conversely, Qwen achieves perfect security (100\%) in both settings and strong privacy (70.3–70.6\%), but its INT4 efficiency regressions (4.775 J, 95.5 ms) undermine the memory benefits of quantization (1.08 GB vs. 2.88 GB FP). Llama-3.2 strikes the most consistent balance, maintaining 80\% jailbreak resistance, the highest privacy (78.2–80.4\%), and moderate efficiency across both settings. Granite-4.1-3b offers an intriguing alternative, achieving perfect jailbreak resistance (100\%) in INT4 and competitive MMLU accuracy (63.9\%), albeit at higher energy and latency costs.

Based on our analysis, several factors contribute to these outcomes:

\begin{itemize}
    \item \textbf{Architectural Safety Prioritization}: Qwen and granite's perfect jailbreak resistance indicates explicit safety alignment in their training objectives, but this may come at the cost of quantization resilience.

    \item \textbf{Quantization Implementation Quality}: Models with well-optimized INT4 implementations (Llama-3.2, gemma-2) maintain efficiency and security, while others (Qwen, granite) suffer overhead.
    
    \item \textbf{Safety Subnetwork Fragility}: Phi-3.5-mini's low jailbreak resistance persists across both FP and INT4, suggesting its safety pathways are inherently sparse and vulnerable.
    
    \item \textbf{Memory vs. Compute Trade-offs}: Models achieving the smallest VRAM footprints under INT4 (Qwen: 1.08 GB, granite: 2.00 GB, gemma-2: 2.08 GB) do not necessarily achieve the lowest energy or latency, highlighting the independent nature of these constraints.
\end{itemize}

\begin{figure}[htbp]
    \centering
    \begin{subfigure}[b]{0.48\textwidth}
        \centering
        \includegraphics[width=\textwidth]{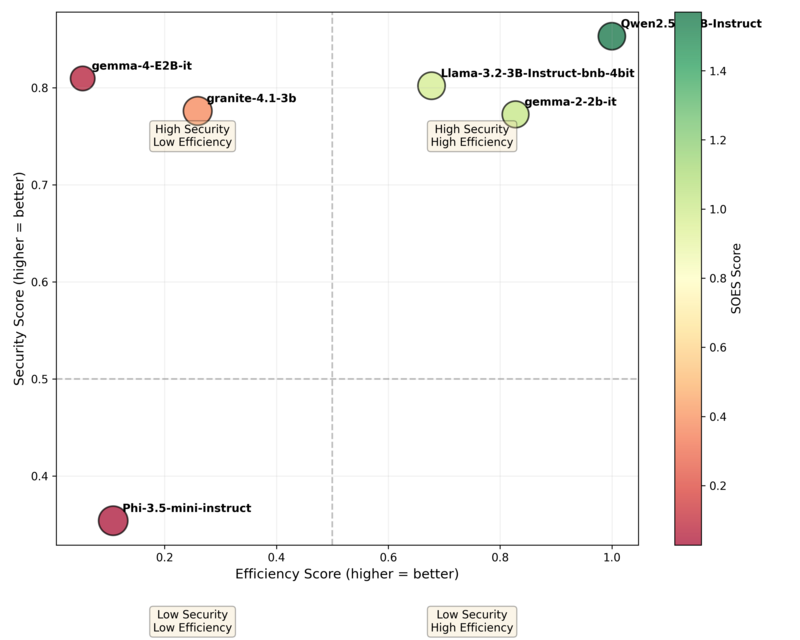}
        \caption{Different LLMs under full precision (FP) inference.}
    \end{subfigure}
    \hfill
    \begin{subfigure}[b]{0.48\textwidth}
        \centering
        \includegraphics[width=\textwidth]{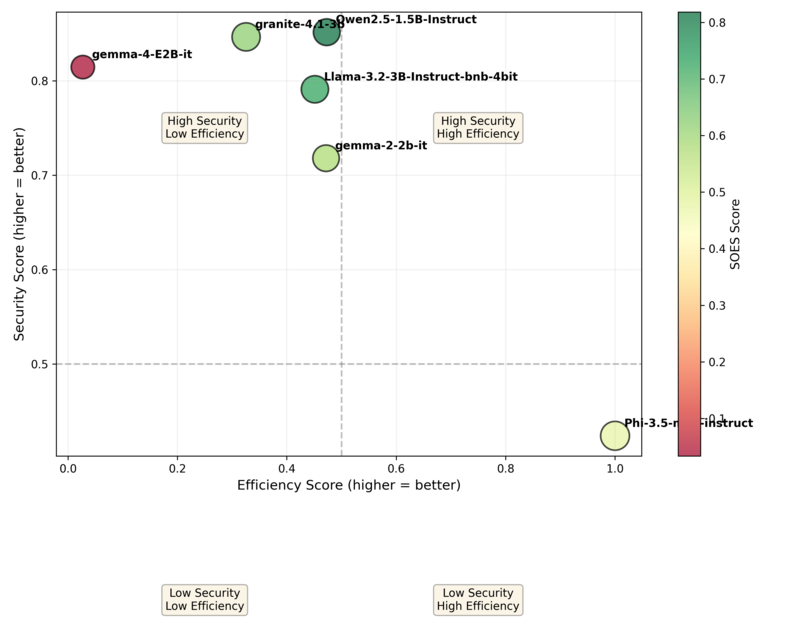}
        \caption{Different LLMs under INT4 quantization.}
    \end{subfigure}
    \caption{The security-efficiency paradox in edge LLMs: Security score vs. Efficiency score under FP and INT4 quantization.}
    \label{fig:sec_eff_paradox_revisit}
\end{figure}

As shown in Figure \ref{fig:sec_eff_paradox_revisit}, the scatter plot of security vs. efficiency reveals distinct trade-off patterns across FP and INT4 configurations. In the full-precision setting, Qwen2.5-1.5B achieves the highest scores in both dimensions, positioning itself near the "High Security, High Efficiency" quadrant, a rare balance that explains its dominant SOES (1.5698). Llama-3.2 and gemma-2-2b-it occupy the upper-middle region, demonstrating strong security with moderate efficiency, while granite-4.1-3b shows competitive efficiency with relatively lower security due to its 90\% jailbreak resistance.

Under INT4 quantization, the landscape shifts notably. Qwen2.5-1.5B maintains its security leadership but experiences a significant efficiency regression, moving toward the "High Security, Low Efficiency" quadrant. Granite-4.1-3b exhibits similar behavior—perfect security (100\%) with reduced efficiency. Conversely, Phi-3.5-mini occupies the "Low Security, High Efficiency" quadrant, reflecting its high MMLU accuracy and low energy consumption but critically low jailbreak resistance (30\%). Llama-3.2 and gemma-2-2b-it strike a middle ground, maintaining moderate security (80\% jailbreak) with balanced efficiency.

This visualization reinforces our central argument: no model simultaneously achieves both high security and high efficiency under INT4 quantization. The FP setting offers a rare exception with Qwen2.5-1.5B, but such configurations are often impractical for edge deployment due to memory constraints (2.88 GB VRAM). Practitioners must navigate this fundamental trade-off based on their specific requirements—prioritizing security, efficiency, or a balanced compromise.

\subsubsection{Practical Considerations and Deployment Guidance}
Based on our comprehensive analysis, we provide the following recommendations for edge LLM deployment. When selecting an edge LLM, practitioners should consider:

\begin{enumerate}
    \item \textbf{Safety-Critical Applications}: Prioritize jailbreak resistance and privacy. Qwen2.5-1.5B is the top choice in both FP (100\% jailbreak, 70.6\% privacy, 2.88 GB) and INT4 (100\%, 70.3\%, 1.08 GB), though INT4 incurs higher energy (4.775 J) and latency (95.5 ms). Granite-4.1-3b (INT4) offers a strong alternative with perfect security and competitive accuracy (63.9\%) at 2.00 GB.

    \item \textbf{Performance-Sensitive Applications}: Phi-3.5-mini achieves the highest MMLU accuracy  and fast INT4 inference (47.0 ms), but its low jailbreak resistance makes it unsuitable for security-sensitive domains. For accuracy without compromising security, granite-4.1-3b (INT4) offers accuracy with perfect jailbreak resistance.

    \item \textbf{Ultra-Constrained Devices}: For devices with stringent memory constraints (e.g., edge microcontrollers, embedded systems), Qwen2.5-1.5B achieves the smallest INT4 footprint at 1.08 GB, followed by granite-4.1-3b, gemma-2-2b-it, and Llama-3.2. While Qwen offers the lowest VRAM usage, its energy and latency regressions under INT4 may offset memory advantages in battery-constrained scenarios. For a more balanced ultra-compact option, Llama-3.2 provides comparable VRAM with superior energy efficiency and lower latency.

    \item \textbf{Battery-Constrained Deployments}: If energy efficiency is critical (e.g., mobile devices, IoT sensors), gemma-2-2b-it demonstrates the lowest FP energy consumption (1.819 J/token) and fastest latency (36.4 ms), while maintaining strong security (90\% jailbreak, 64.5\% privacy). Under INT4, Phi-3.5-mini becomes the most energy-efficient at 2.349 J/token, followed by gemma-2-2b-it (3.446 J) and Llama-3.2 (3.506 J). Practitioners must weigh energy savings against the security compromises of Phi-3.5-mini.
\end{enumerate}

Finally, these findings underscore the need for co-designing optimization techniques and safety alignments, rather than treating them as independent objectives. Future work should explore safety-aware quantization \cite{lee2025quantization} and architectural innovations that preserve robustness under compression.

\section{Conclusion}
Deploying LLMs to the edge mitigates cloud-centric privacy risks but introduces distinct attack vectors rooted in the Security-Efficiency Paradox. Aggressive optimizations such as quantization, pruning, partitioning, and PEFT, that break through the Memory, Quadratic, and Compute Walls simultaneously weaken safety alignment, enable reconstruction attacks, and expose membership information. We have shown that these risks are not incidental but structural, arising from the mathematical nature of low-precision arithmetic, sparse subnetworks, and low-rank adaptation. To address this, we proposed a Three-Wall Constraint Model to quantify when unsafe optimizations become unavoidable, and the Secure Operational Efficiency Score (SOES) to select models that balance utility, security, and efficiency. Our framework moves beyond isolated defenses toward co-designed hardware-aware algorithms and verifiable safety alignments. Based on our findings, we suggest that future work should focus on a) safety-aware compression objectives, b) hardware-level security primitives (e.g., encrypted memory regions for safety-critical weights), and c) real-time monitoring of alignment drift in continuously adapting edge models.

\section*{Acknowledgments}
The authors partially used AI tools to help create selected illustrative figures and experimental code in this article. All technical content, interpretations, and conclusions were reviewed and verified by the authors.

\bibliographystyle{unsrt} 
\bibliography{references} 

\end{document}